%% file: mem_v4.tex
\documentclass[%
  reprint,
showpacs,preprintnumbers,
  amsmath,amssymb,
  aps,
  floatfix,
]{revtex4-1}

\usepackage{graphicx}% Include figure files
\usepackage{dcolumn}% Align table columns on decimal point
\usepackage{bm}% bold math
\usepackage{hyperref}% add hypertext capabilities

\makeatletter
\newsavebox{\@brx}
\newcommand{\llangle}[1][]{\savebox{\@brx}{\(\m@th{#1\langle}\)}%
  \mathopen{\copy\@brx\mkern2mu\kern-0.9\wd\@brx\usebox{\@brx}}}
\newcommand{\rrangle}[1][]{\savebox{\@brx}{\(\m@th{#1\rangle}\)}%
  \mathclose{\copy\@brx\mkern2mu\kern-0.9\wd\@brx\usebox{\@brx}}}
\makeatother

\begin{document}

\preprint{J-PARC-TH-0070}

\title{
In-medium dispersion relations of charmonia
studied by maximum entropy method}

\author{Atsuro Ikeda}
\email[]{a-ikeda@kern.phys.sci.osaka-u.ac.jp}
\affiliation{Department of Physics, Osaka University, Toyonaka, Osaka 560-0043, Japan}
\author{Masayuki Asakawa}
\email[]{yuki@phys.sci.osaka-u.ac.jp}
\affiliation{Department of Physics, Osaka University, Toyonaka, Osaka 560-0043, Japan}
\author{Masakiyo Kitazawa}
\email[]{kitazawa@phys.sci.osaka-u.ac.jp}
\affiliation{Department of Physics, Osaka University, Toyonaka, Osaka 560-0043, Japan}
\affiliation{J-PARC Branch, KEK Theory Center,
  Institute of Particle and Nuclear Studies, KEK,
  203-1, Shirakata, Tokai, Ibaraki, 319-1106, Japan }

\date{\today}% It is always \today, today,

\begin{abstract}
  We study in-medium spectral properties of charmonia 
  in the vector and pseudoscalar channels at nonzero momenta on
  quenched lattices, especially focusing on their dispersion relation
  and weight of the peak.
  We measure the lattice Euclidean correlation functions with nonzero 
  momenta on the anisotropic quenched lattices 
  and study the spectral functions with the maximum entropy method.
  The dispersion relations of charmonia and the momentum dependence 
  of the weight of the peak are analyzed with the maximum entropy method
  together with the errors 
  estimated probabilistically in this method.
  We find significant increase of the masses of charmonia in medium.
  It is also found that the functional form  of the charmonium dispersion
  relations is not changed from that in the vacuum within the error 
  even at $T\simeq1.6T_{\rm c}$ for all the channels we analyzed.
\end{abstract}

\maketitle

%\tableofcontents

\section{introduction}
\label{sec:intro}

Understanding properties of heavy quarkonia in hot medium near and above 
the critical temperature $T_{\rm c}$ of the deconfinement phase 
transition is one of the important subjects in relativistic heavy 
ion collisions \cite{brambilla_heavy_2011,andronic_heavy-flavour_2016}.
It is believed that, 
by understanding the stabilities of heavy quarkonia, 
their yields in heavy ion collisions can be used as 
a signal of the formation of the quark-gluon plasma 
\cite{matsui_j/_1986}.
Because of their heavy mass, heavy quarks
(charm and bottom quarks and their antiparticles) also serve as unique 
theoretical and experimental probes to diagnose the properties
of the hot medium, such as the heavy-quark potential \cite{hashimoto_mass_1986} 
and the transport properties \cite{akamatsu_heavy_2009}.
Experimental progress in this field, such as the isolation of 
charm and bottom quarks using the silicon vertex tracker 
\cite{phenix_collaboration_measurement_2009},
will enrich the study of heavy quarks further.

The hot matter created by heavy ion collisions with temperature ($T$)
near $T_{\rm c}$ is believed to be a strongly interacting system.
Lattice QCD is a powerful method to investigate such a region of QCD 
at which nonperturbative effects will play a crucial role.
The study of the properties of charmonia at finite temperature 
is one of the longstanding subjects on the lattice 
\cite{asakawa_j/_2004,datta_behavior_2004,umeda_charmonium_2004,
datta_meson_2005,jakovac_quarkonium_2007, umeda_constant_2007,
 whot-qcd_collaboration_charmonium_2011,ding_charmonium_2012-1,
borsanyi_charmonium_2014}. 
In lattice QCD numerical simulations, which 
rely on the imaginary time formalism, however, one cannot analyze 
dynamical properties encoded in spectral functions directly.
Instead, only Euclidean correlation functions are calculable on the lattice.
To obtain the spectral functions from 
the Euclidean correlators, one has to take an analytic
continuation from imaginary time to real time.
The maximum entropy method (MEM) is
a useful method to perform this analytic continuation on
the basis of probability theory.
\cite{jarrell_bayesian_1996,asakawa_maximum_2001}.
The studies of charmonium spectral functions with MEM 
qualitatively agree with each other in that the charmonia 
in the vector and pseudoscalar channels survive up to around 
$T \simeq 1.5T_{\rm c}$
\cite{asakawa_j/_2004,datta_behavior_2004,datta_meson_2005,
borsanyi_charmonium_2014}.

In the previous studies on heavy quarkonia on the lattice,
the analyses have been performed only for zero momentum 
with a few exceptions
\cite{datta_charmonia_2004, nonaka_charmonium_2011,
oktay_momentum-dependence_2010, aarts_s_2013,ding_momentum_2013}.
The spectral function with zero momentum represents
the spectral properties of a charmonium at rest in medium.
On the other hand, charmonia in the hot medium created by heavy 
ion collisions typically have nonnegligible velocity against the rest 
frame of the medium because 
charmonia generated by hard processes 
in the early stage can have large momentum.
The finiteness of the velocity of charmonia may modify 
their properties, such as the stability
\cite{liu_antichar21sitter/conformal-field-theory_2007}
and the dispersion relation, i.e. 
the momentum dependence of energy.
Here it is worthwhile to note that such modifications of the 
dispersion relations in medium are suggested in various systems
\cite{bellac_thermal_2000,shuryak_physics_1990,pisarski_propagation_1996}, 
and such a modification can give rise to novel phenomena such as 
van Hove singularity 
\cite{van_hove_occurrence_1953,braaten_production_1990,
kitazawa_possible_2014,kim_dilepton_2015}.
It thus is interesting to explore the momentum 
dependence of spectral functions of charmonia and 
their dispersion relations near and above $T_{\rm c}$ 
with the first principle simulation on the lattice.
The purpose of the present study is to perform a quantitative 
study on the momentum dependence of charmonium spectral functions 
on the lattice with MEM.

In this study we explore the properties of charmonia in the 
vector and pseudoscalar channels, corresponding to
$J/\psi$ and $\eta_c$, respectively,
at nonzero momenta on anisotropic quenched lattices.
In addition to the standard analysis of the spectral functions in MEM, 
we study the dispersion relations and the momentum dependence of
the spectral weights of 
the $J/\psi$ and $\eta_c$ peaks on the basis of MEM.
To perform the measurement of the dispersion relation with 
a quantitative error analysis in MEM, 
we analyze the center of weight of the peak in the spectral function.
As we will see later, this quantity is identical to the peak position 
for sufficiently narrow peaks, but error analysis can be carried out
in MEM. 
Similarly, we analyze the weight of the peak, which corresponds to 
the residue of the peak,
with the error 
analysis. 
For the vector channel, the transverse and longitudinal 
components are investigated separately in the analysis.

We find that the masses of $J/\psi$ and $\eta_c$ defined by
the dispersion relation at zero momentum 
show significant increase as $T$ is raised.
It is also found that the dispersion relation of charmonia
continues to take the Lorentz covariant form, i.e. the same form as
in the vacuum, even well above 
$T_{\rm c}$ within the error.
Our numerical analysis also suggests that the weight of the peak at finite
temperature
does not have momentum dependence within the error.

This paper is organized as follows.
In Sec.~\ref{sec:SPF}, we introduce 
the spectral function and summarize its properties. 
We then discuss MEM in Sec.~\ref{sec:mem}.
The error estimate in this method and the quantities
corresponding to the dispersion relation and spectral weight of 
charmonia are discussed in this section.
In Sec.~\ref{sec:lattice}, we show our lattice set up.
We then discuss the numerical results in Sec.~\ref{sec:result}.
Sec.~\ref{sec:conclusion} is devoted to conclusion.

\section{Euclidean correlator and spectral function}
\label{sec:SPF}

Dynamical properties of charmonia are encoded in 
the Euclidean correlators% with momentum $\vec{p}$
\begin{align}
  \label{eq:operator}
  G^{lm}(\tau,\vec{p})=\int d^3xe^{i\vec{p}\cdot\vec{x}} \left\langle J^l(\tau,
  \vec{x})J^{m\dagger}(0,\vec{0})\right\rangle ,
\end{align}
where the imaginary time $\tau$ is restricted to the interval
$ 0\leq\tau< 1/T $ and 
$J^l(\tau,\vec{x})= \bar{c}(\tau,\vec{x})i\gamma^l c(\tau,\vec{x})$ 
is the local interpolating operator in the Heisenberg representation
with the charm quark field 
$c(\tau,\vec{x})$ with $l=0,1,2,$ and $3$ for the vector channel 
and $l=5$ for the pseudoscalar channel. 
The spectral function $A^{lm}(\omega,\vec{p})$ is defined as the
imaginary part of the
retarded correlator $G^{lm}_{\rm R}(\omega,\vec{p})
$ divided by $\pi$, 
\begin{align}
  \label{eq:spf1}
  A^{lm}(\omega,\vec{p})
  &= 
  \frac{1}{\pi}{\rm Im} \left[ G^{lm}_{\rm R}(\omega,\vec{p})
  \right],
\end{align}
where
\begin{align}
  \label{eq:spf2}
   G^{lm}_{\rm R}(\omega,\vec{p})&=
 \int_{-\infty}^{\infty}dt\int d^3 \vec{x} e^{-i\vec{p}\cdot\vec{x}+i\omega
 t}G^{lm}_{\rm R}(t,\vec{x}), \\
G^{lm}_{\rm R}(t,\vec{x})&=
    i \theta(t) \langle [J^l(t,\vec{x}),J^{m\dagger}(0,\vec{0})]\rangle.
  \label{eq:spf3}
\end{align}
The diagonal components of the spectral functions
$A^{ll}(\omega,\vec{p})$
are related to Eq.~(\ref{eq:operator}) 
by the Laplace-like transformation as 
\begin{align}
  G^{ll}(\tau,\vec{p})
  = \int_0^\infty {K}(\tau,\omega){A^{ll}}{(\omega,\vec{p})} d\omega ,
  \label{eq:correlation_function}
\end{align}
with 
\begin{align}
  K(\tau,\omega)=\frac{ e^{-\tau\omega}+e^{-(1/T-\tau)\omega}} 
  {1-e^{-\omega/T}}.
  \label{eq:kernel}
\end{align}
In the following, we represent the diagonal components of the 
spectral functions as 
\begin{align}
A^l(\omega,\vec{p}) = A^{ll}(\omega,\vec{p}).
\end{align}

In the vacuum, as a consequence of Lorentz invariance and charge 
conservation, the vector spectral function can be represented as
\begin{align}
  A^{\mu\nu}(\omega,\vec{p}) =\left(\frac{p^\mu p^\nu}{P^2}-g^{\mu\nu}\right)
  A_{\rm V}(P^2),
  \label{eq:vector_spectral_vacuum}
\end{align}
with $\mu,\nu=0,1,2,$ and $3$ and $P^2 = \omega^2 - |\vec{p}|^2$.
When there is a bound state which couples to $J^l$,
the corresponding spectral function
$A_{\rm V}(\omega,p)=A_{\rm V}(P^2)$
or $A_{\rm PS}(\omega,p)=A^5(\omega,\vec{p})$ has
a peak structure around $\omega\simeq \pm E({p})$, where
$E({p})$ is the dispersion relation of the bound state with $p=|\vec{p}|$.
The peak structure is approximately be given by a delta function,
\begin{align}
  Z\delta(\omega^2-E(p)^2)
  =\frac{Z}{2E(p)}\delta\left(\omega-E(p)\right),
  \label{eq:peak_structure}
\end{align}
where the right hand side
represents the peak at $\omega>0$, 
and $Z>0$ is the residue.
Because of Lorentz invariance, $E(p)$ in the vacuum is given by
\begin{equation}
  E(p)=\sqrt{m^2+p^2},
  \label{eq:dispersion_vacuum}
\end{equation}
where $m$ is the mass of the bound state.
It is also shown from Lorentz invariance that 
$Z$ in Eq.~(\ref{eq:peak_structure}) does not have momentum dependence.

The property of the bound state 
peak
in Eqs.~(\ref{eq:peak_structure}) and (\ref{eq:dispersion_vacuum})
is modified at finite temperature. 
First, 
the width of the peak becomes larger and 
the delta function in Eq.~(\ref{eq:peak_structure}) 
is
replaced by a smooth function with a peak.
Second, because Lorentz invariance is lost in medium,
$Z$ can depend on momentum.
The dispersion relation $E(p)$ can also be modified from the
Lorentz covariant form Eq.~(\ref{eq:dispersion_vacuum}).

At finite temperature, 
$A^{\mu\nu}(\omega,\vec{p})$ 
in Eq.~(\ref{eq:vector_spectral_vacuum}) is decomposed into 
the transverse and longitudinal components as
\cite{kapusta_finite-temperature_2006}
\begin{equation}
  A^{\mu\nu}(\omega,\vec{p}) = P_{\rm T}^{\mu\nu} A_{\rm T}(\omega,p)
  +P_{\rm L}^{\mu\nu} A_{\rm L}(\omega,p) 
  \label{eq:decomposition}, 
\end{equation}
where the projection operators onto the transverse and longitudinal
components, $P_{\rm T}$ and $P_{\rm L}$, respectively, are defined as 
\begin{align}
  &P_{\rm T}^{00}=P_{\rm T}^{0i}=P_{\rm T}^{i0}=0,\\
  &P_{\rm T}^{ij}=\delta^{ij}-p^ip^j/p^2, \\
  &P_{\rm L}^{\mu\nu}=p^\mu p^\nu/P^2 -g^{\mu\nu} -P_{\rm T}^{\mu\nu},
\end{align}
with $i,j=1,2,$ and $3$.
The transverse and longitudinal spectral functions
$A_{\rm T}(\omega,p)$ and $A_{\rm L}(\omega,p)$ are identical in the vacuum,
$A_{\rm T}(\omega,p) = A_{\rm L}(\omega,p) =A_{\rm V}(\omega,p)$, from
Eq.~(\ref{eq:vector_spectral_vacuum}).
When the momentum is taken as $\vec{p}=(p,0,0)$,
$A_{\rm T}(\omega,p)$ and $A_{\rm L}(\omega,p)$ are related to 
$A^i(\omega,\vec{p})$ as
\begin{align}
A_{\rm T}(\omega,p) &= \frac12 \left( A^2(\omega,\vec{p}) 
+ A^3(\omega,\vec{p}) \right),
\label{eq:A_T}
\\
A_{\rm L}(\omega,p) &=  \frac{\omega^2-{p}^2}{\omega^2} A^1(\omega,\vec{p}) .
\label{eq:A_L}
\end{align}

From the general property of the spectral function,
$A^l(\omega,\vec{p})$ are semi-positive 
for $\omega>0$ \cite{kapusta_finite-temperature_2006}.
The semi-positivity of $A_{\rm T}(\omega,p)$ is then 
guaranteed from Eq.~(\ref{eq:A_T}).
On the other hand, Eq.~(\ref{eq:A_L}) shows that 
$A_{\rm L}(\omega,p)$ is semi-negative in the space-like region 
$0<\omega<p$.

\section{maximum entropy method}
\label{sec:mem}
In this section we give a briefly review of MEM and show
how the dispersion relations and spectral weights of charmonia
are estimated in our analysis using MEM.

To obtain the spectral function from the lattice Euclidean correlator, 
we have to take 
the inverse transformation of Eq.~(\ref{eq:correlation_function}).
MEM \cite{jarrell_bayesian_1996, asakawa_maximum_2001}
is a method to infer the most probable image of the spectral function 
from a limited number of data points for a Euclidean correlator 
on the basis of Bayes' theorem.

In the analysis of a spectral function $A(\omega)$ %that gives
corresponding to 
a Euclidean correlator $G(\tau)$ obtained in a 
Monte Carlo simulation with Eq.~(\ref{eq:correlation_function}),
the most important quantity is the $\chi$-square, 
\begin{align}
  \chi^2=
  \sum_{i,j} \left(G(\tau_i)-G_A(\tau_i) \right)
  C_{ij}^{-1}\left(G(\tau_j)-G_A(\tau_j)\right) ,
  \label{eq:chi_square} %\\
\end{align}
where the correlation between different temporal points $\tau_i$ 
is encoded in the covariance matrix $C_{ij}$,  
$i$ and $j$ run over discrete temporal points and
$G_A(\tau_i)$ is the correlator defined by 
Eq.~(\ref{eq:correlation_function}) from the spectral function
$A(\omega)$.

In the standard least-square method, $A(\omega)$ is 
determined so as to minimize Eq.~(\ref{eq:chi_square}).
Because the number of the degrees of the freedom of the continuous 
function $A(\omega)$ is larger than the one of the discrete data 
for $G(\tau)$, however, the minimum of $\chi^2$ is heavily 
degenerating. To choose one, some ansatz to constrain 
the functional form of $A(\omega)$ is required.

In order to remove this degeneracy, 
MEM introduces a prior probability represented by
the Shannon-Jaynes entropy \cite{bryan_maximum_1990},
\begin{align}
  S=\int_0^{\infty} \left[ A(\omega)-m(\omega) -A(\omega)\log\left( 
    \frac{A(\omega)}{m(\omega)}
  \right) \right]d\omega ,
  \label{eq:shannon-jaynes_entropy}
\end{align}
where 
the default model $m(\omega)$ expresses prior knowledge.
From Bayes' theorem, it is obtained that 
the conditional probability of having $A(\omega)$ 
from $G(\tau)$ and the prior knowledge is proportional to 
$P(A,\alpha) = \exp[Q(A,\alpha)]$ \cite{asakawa_maximum_2001}, where
\begin{align}
  Q(A,\alpha) = \alpha S(A) - \frac12 \chi^2(A).
  \label{eq:Q}
\end{align}
The parameter $\alpha$ controls the relative weight between $\chi^2$ and $S$.
It is known that the spectral image that maximizes $P(A,\alpha)$
for a given $\alpha$ is unique if it exists \cite{asakawa_maximum_2001}.
The final output image $A_{\rm out}(\omega)$ is obtained 
by integrating $A(\omega)$ with a weight $P(A,\alpha)$ 
over $\alpha$ and $A$ space as
\begin{align}
  A_{\mathrm{out}}(\omega) =
  \llangle A(\omega) \rrangle , 
  \label{eq:Aout1}
\end{align}
where 
\begin{align}
  \llangle {\cal O} \rrangle 
=   \frac{1}{Z_{\rm P}} \int d\alpha \int \left[ dA \right]  
P(A,\alpha) {\cal O},
\label{eq:<<O>>}
\end{align}
is the average over the plausibility
$P(A,\alpha)$ with 
$Z_{\rm P}\equiv \int d\alpha\int [dA] P(A,\alpha)$.
Here, the measure $[dA]$ is defined as
\begin{align}
  [dA] \equiv \lim_{N_\omega \to
  \infty}\prod_{l=1}^{N_\omega}\frac{dA_l}{\sqrt{A_l}},
  \label{eq:measure_A}
\end{align}
with the discretized spectral function 
$A_l= A(\omega_l)$ with discrete $\omega$ values $\omega_l$
\cite{asakawa_maximum_2001}.
When $P(A,\alpha)$ is sharply peaked around 
$A_\alpha(\omega)$, Eq.~(\ref{eq:Aout1}) is well approximated as
\begin{align}
  A_{\mathrm{out}}(\omega) \simeq \frac{1}{Z_{\rm P}}
  \int d\alpha A_\alpha(\omega)P(\alpha),
  \label{eq:Aout2}
\end{align}
where
\begin{align}
P(\alpha)\equiv \int [dA] P(A,\alpha).
\label{eq:P(a)}
\end{align}

\subsection{Error analysis}
\label{sec:error}

A characteristic of MEM is that this method enables us to estimate
the error of quantities given by the integral of a function of 
$A_{\rm out}(\omega)$ quantitatively.

Let us consider a quantity
given by the weighted integral of $A(\omega)$
with a weight function $f(\omega)$ and an interval 
$I=[\omega_{\rm min},\omega_{\rm max}]$,
\begin{align}
W = \int_I f(\omega)  A(\omega) d\omega .
\label{eq:W}
\end{align}
In MEM, the average of $W$ is estimated as 
\begin{align}
  \langle W \rangle
  =  
  \llangle[\bigg] \int_{I} d\omega f(\omega) A(\omega) \rrangle[\bigg] ,
\label{eq:average_A}
\end{align}
and the error of $\langle W\rangle$ is given by the 
variance of $W$ in $P(A,\alpha)$ space as 
\begin{align}
\Delta W = \sqrt{ \llangle (\delta W)^2 \rrangle },
\label{eq:error_A}
\end{align}
where $\delta W = W - \llangle W \rrangle$.

Typically, the magnitude of the error estimated in this way 
becomes larger as the interval $I$ becomes narrower. 
In particular, when one makes the error estimate of 
$A_\mathrm{out}(\omega)$ at a given $\omega$
with $f(\omega')=\delta(\omega'-\omega)$, 
one obtains a huge error 
$\Delta A_\mathrm{out}(\omega)\gg A_\mathrm{out}(\omega)$.
This means that the functional form of $A_\mathrm{out}(\omega)$ 
itself does not have quantitative meaning.
For example, it does not make sense to distinguish whether
the functional form of a peak structure in $A_\mathrm{out}(\omega)$ 
is Gaussian or Lorentzian in MEM.
The values of the position and width of the peak do not have 
statistically relevant meanings, either.
In order to obtain a moderate value of the error, the interval
$I$ has to be chosen sufficiently large.
This limitation of the analysis is associated with 
reconstructing apparently more information than the 
original one.
Even if the correlators $G(\tau_i)$ for discrete $\tau_i$'s are 
determined with an infinitesimal statistical error, 
the reconstructed image $A_{\rm out}(\omega)$ still have error.
This is because the error in MEM includes intrinsic
one associated with the introduction of the entropy,
in addition to statistical one.
Thus, for instance,
it is not sufficient to estimate the error in the result with
the Jackknife methods, which takes account of only the statistical
error. 
The error analysis with
Eq.~(\ref{eq:error_A}) is essential and 
absolutely necessary \cite{asakawa_maximum_2001}.

\subsection{Dispersion relation and residue}
\label{sec:ZMEM}

In this study, we focus on the dispersion relation and 
the momentum dependence of the spectral weight of the $J/\psi$ and $\eta_c$.
To study these quantities with error estimates in MEM, 
we have to represent $E(p)$ and $Z$ in Eq.~(\ref{eq:peak_structure}) 
in the form in Eq.~(\ref{eq:W}).

For such a quantity corresponding to the residue $Z$, 
we consider 
\begin{align}
\bar{Z}(p) = \int_I d\omega 2\omega A(\omega, p),
\label{eq:barZ}
\end{align}
for a peak in a spectral function $A(\omega, p)$,
where $I$ is the interval of $\omega$ which covers the peak
structure.
By substituting Eq.~(\ref{eq:peak_structure}) into Eq.~(\ref{eq:barZ})
one easily finds that for the delta function 
Eq.~(\ref{eq:peak_structure}) we have $\bar{Z}(p)=Z$.
When the interval $I$ does not include other structures in $A(\omega)$,
therefore, $\bar{Z}(p)$ corresponds to $Z$.
Note that 
$\bar{Z}(p)$ defined by Eq.~(\ref{eq:barZ}) is
meaningful only for well isolated 
peaks for which such a choice of $I$ is possible.
Since Eq.~(\ref{eq:barZ}) has the form given in Eq.~(\ref{eq:W}),
$\bar{Z}(p)$ is a quantity which can be estimated in MEM 
with error.

Next, to analyze the dispersion relation in MEM, 
we consider the center of the weight of a peak of 
the dimensionless spectrum $A(\omega)/\omega^2$, which is given by
\begin{align}
  \bar{E}(p)= \frac{ \int_I d\omega \omega ( A(\omega,p)/\omega^2 ) }
{ \int_I d\omega (A(\omega,p)/\omega^2 ) }.
\label{eq:barE}
\end{align}
By substituting Eq.~(\ref{eq:peak_structure}) into Eq.~(\ref{eq:barE}),
it is again checked that $\bar{E}(p)=E(p)$ for this case.
In practical analysis, we calculate Eq.~(\ref{eq:barE}) as 
\begin{align}
  \bar{E}(p)=
  \frac{\llangle \int_I d\omega \omega ( A(\omega,p)/\omega^2 ) \rrangle}
       { \llangle \int_I d\omega (A(\omega,p)/\omega^2 ) \rrangle},
  \label{eq:center_of_weight}
\end{align}
in order to perform the analysis with the saddle point 
approximation for $P(A,\alpha)$ \cite{asakawa_maximum_2001}.
In the error analysis for Eq.~(\ref{eq:center_of_weight}),
we
take account of the correlation between the numerator and 
denominator using 
the general formula of error propagation.
Because the numerator and denominator are positively correlated,
the inclusion of this correlation leads to the suppression of the error 
of Eq.~(\ref{eq:center_of_weight}).

In the above discussions, the definitions of $\bar{Z}(p)$ 
and $\bar{E}(p)$ depend on the energy interval $I$.
Because the choice of the interval has an arbitrariness,
in the analyses of these quantities 
one has to check the dependence of $\bar{Z}(p)$ and $\bar{E}(p)$ on 
the interval $I$ by varying it in a moderate range.
This analysis will be performed in Sec.~\ref{sec:residue}.
As we will see there, the peaks corresponding to the $J/\psi$ and 
$\eta_c$ analyzed in this study are well isolated and our results 
on $\bar{Z}(p)$, $\bar{E}(p)$, and their errors are 
insensitive to the choice of $I$.

\section{Simulation set up}
\label{sec:lattice}
  \begin{table}[tb]
    \centering
    \caption{\label{tab:lattice_setup}Lattice simulation parameters.}
    \begin{tabular}{c|ccccc}
      \hline
      $T/T_{\mathrm{c} }$ & $N_\tau$ & $N_\sigma$ &
      $N_{\mathrm{conf}}$\\ \hline
      1.70 & 44 & 64 & 700\\
      1.62 & 46 & 64 & 500\\
      1.56 & 48 & 64 & 500\\
      1.49 & 50 & 64 & 500\\
      1.38 & 54 & 64 & 500\\
      0.78 & 96 & 64 & 500\\
      \hline
    \end{tabular}
  \end{table}
In this study, we measure the momentum dependence of charmonium 
correlation functions Eq.~(\ref{eq:operator})
in the vector and pseudoscalar channels on 
quenched anisotropic lattices with the standard Wilson gauge action
and Wilson fermion.
The simulation parameters are $\beta=7.0$, 
the bare anisotropy $\xi_0=3.5$, the spatial hopping parameter 
$\kappa_\sigma=0.08285$, and the fermion anisotropy $\gamma_F=3.476$ 
\cite{asakawa_j/_2004}.
The anisotropy $\xi=a_\sigma/a_\tau$ is $4$ with the lattice spacings along 
spatial and temporal directions, $a_\sigma$ and $a_\tau$, respectively.
The temporal lattice spacing in physical unit is 
$a_\tau=0.00975$~fm \cite{asakawa_j/_2004}.
In Table.~\ref{tab:lattice_setup}, we summarize the lattice parameters,
the lattice volumes $N_\sigma^3\times N_\tau$, the temperature $T$
in the unit of $T_{\rm c}$ \cite{asakawa_j/_2004}, 
and the number of configurations $N_{\rm conf}$.
We fix the lattice size for spatial direction to $N_\sigma=64$.
With the aid of the large anisotropy, our lattice has 
a large spatial volume; in physical unit 
the spatial length is $L_\sigma\simeq2.5$~fm.
The aspect ratio $\xi N_\sigma/N_\tau$ is $5.1$ for $N_\tau=50$.
The large spatial extent enables 
a detailed study of the momentum dependence of the quantities on the lattice.
With a periodic boundary condition along spatial direction,
the momentum of bosons on the lattice is discretized as
\begin{align}
p_i=\frac{2}{a_\sigma}\sin\left(\frac{\pi\hat{p}_i}{N_\sigma}\right), 
\label{MomLat}
\end{align}
with integer $\hat{p}_i$ in $-N_\sigma/2 < \hat{p}_i \le N_\sigma/2$.
In the analysis of Euclidean correlators we take the momentum 
along $1$ direction, i.e. $\vec{p}=(p,0,0)$.
The largest lattice with $N_\tau=96$ and $T/T_{\rm c}=0.78$
is regarded as the vacuum one,
in which the medium effects are well suppressed.

To improve statistics, 
we have measured the correlation functions eight times on each 
gauge configuration with different positions of the point source.
The eight sources are located on two timeslices separated by $N_\tau/2$
with four sources in each timeslice with maximal separation.
The correlation function on a configuration is then defined 
by the average of these eight measurements.
We have checked that the statistics improves about $\sqrt{8}$ times
with this treatment, suggesting that the correlation between 
eight measurements is well suppressed.
Because of this treatment, our numerical analyses have an advantage
in the statistics of the correlation function compared to the previous studies.
The number of $N_\tau$ is also large in our analyses thanks to
the large anisotropy compared to the other studies of spectral
functions in MEM.
These characteristics of our analysis enable us to obtain 
numerical results with high resolution.

In the MEM analysis of spectral functions,
we use a default model $m(\omega)=m_0\omega^2+m_1T\omega$
\cite{aarts_spectral_2007},
where $m_0=1.15$ for the
pseudoscalar channel and $m_0=0.40$ for the vector channel 
\cite{asakawa_j/_2004}.
We analyzed the default model dependence of the reconstructed
spectral function by changing $m_0$ and $m_1$:
$m_0$ is varied in the $50\%$ range from the above values, 
while the dependence on $m_1$ is also checked 
in the range $0\le m_1 \le 2.0$.
We found that the dependence 
on $m_0$ and $m_1$ is well suppressed in this range
near the peak of the $J/\psi$ and $\eta_c$; the change of the spectral 
image caused by the variation of $m_0$ and $m_1$ in these ranges
around the charmonium peak is less than a few percent.
In the following analyses we thus set $m_1=0$.
For nonzero $p$ and large $\omega$, it may be better to replace 
the default model by $m(\omega)=m_0(\omega^2-p^2)$ 
as suggested from Lorentz invariance.
We have performed MEM analysis with %the form of the 
a default model
\begin{align}
m(\omega) = 
\begin{cases}
m_0(\omega^2-p^2) & \omega^2 \ge p^2+\epsilon \\
m_0 \epsilon & \omega^2<p^2+\epsilon
\end{cases},
\end{align}
with several choices of small parameter $\epsilon$.
We, however, found that the default model dependence of our results
is well suppressed again around the peak of the $J/\psi$ and $\eta_c$,
and the following numerical results hardly change.

For the analysis of $A_{\rm T}(\omega,p)$ and $A_{\rm PS}(\omega,p)$,
we reconstruct them from the corresponding correlators
$G_{\rm T}(\tau,p) = ( G^2(\tau,\vec{p}) + G^3(\tau,\vec{p}) )/2$ and 
$G_{\rm PS}(\tau,p) = G^5(\tau,\vec{p})$.
On the other hand, $A_{\rm L}(\omega,p)$ can become negative and 
one cannot apply MEM to this channel directly.
We thus analyze $A^1(\omega,\vec{p})$ from $G^1(\tau,\vec{p})$ 
with MEM and obtain $A_{\rm L}(\omega,p)$ using Eq.~(\ref{eq:A_L}).

For sets of gauge configurations with $N_\tau=48$ and $54$, 
we observed that the reconstructed spectral images obtained 
in MEM analysis behave in an unreasonable way in some channels.
We found that  
the error of the spectral functions tends to become large when
such behaviors are observed.
For example, although the spectral functions $A^i(\omega,\vec{p})$ 
with $\vec{p}=0$ should be degenerated for $i=1,2,$ and $3$ because of
rotational invariance, for $N_\tau=48$ and $54$, we observed 
that the reconstructed images of $A^i(\omega,\vec{p})$ behave
in qualitatively different ways.
It is also found that $P(\alpha)$ as a function of $\alpha$ 
has rapid changes for some values of $\alpha$ when such a 
pathological behavior is observed in MEM analysis.
We have checked that these results do not come from the numerical
resolution in our MEM algorithm by changing the numerical precision of 
our code.
We have also checked that they do not depend on the choice 
of the default model.
This problem is discussed in detail in Appendix~\ref{sec:pathology}.
In this study, we simply exclude $N_\tau=48$ and $54$ 
in the following discussion and concentrate on 
$N_\tau=44,~46,~50,$ and $96$, which do not show such behaviors.

In the analysis of the dispersion relation Eq.~(\ref{eq:center_of_weight})
and the weight of the peak Eq.~(\ref{eq:barZ}),
the interval $I=[\omega_{\rm min},\omega_{\rm max}]$ has to be chosen 
appropriately.
In this study, we set $\omega_{\rm min}=3$~GeV, while 
for the upper bound $\omega_{\rm max}$ we use the value of $\omega$ 
at which the spectral function takes the first local minimum 
on the right of the peak corresponding to the $J/\psi$ or $\eta_c$.
We found that our numerical results for $\bar{Z}(p)$ and 
$\bar{E}(p)$, as well as their errors, are insensitive to the choice of the lower bound $\omega_{\rm min}$;
for example, these quantities do not change within the numerical 
precision even if $\omega_{\rm min}$ is set to $2$~GeV.
Our numerical analysis suggests that the results of 
$\bar{Z}(p)$ and $\bar{E}(p)$ hardly 
change for a variation of the lower and/or upper limits of $I$ in 
the range where the reconstructed image takes a small value.
The dependence of our results on the choice of $\omega_{\rm max}$
will be discussed in Sec.~\ref{sec:residue}.

\section{Numerical results}
\label{sec:result}
\subsection{Correlation function}
  \begin{figure}[tb]
    \centering
    \includegraphics[width=0.49\textwidth]{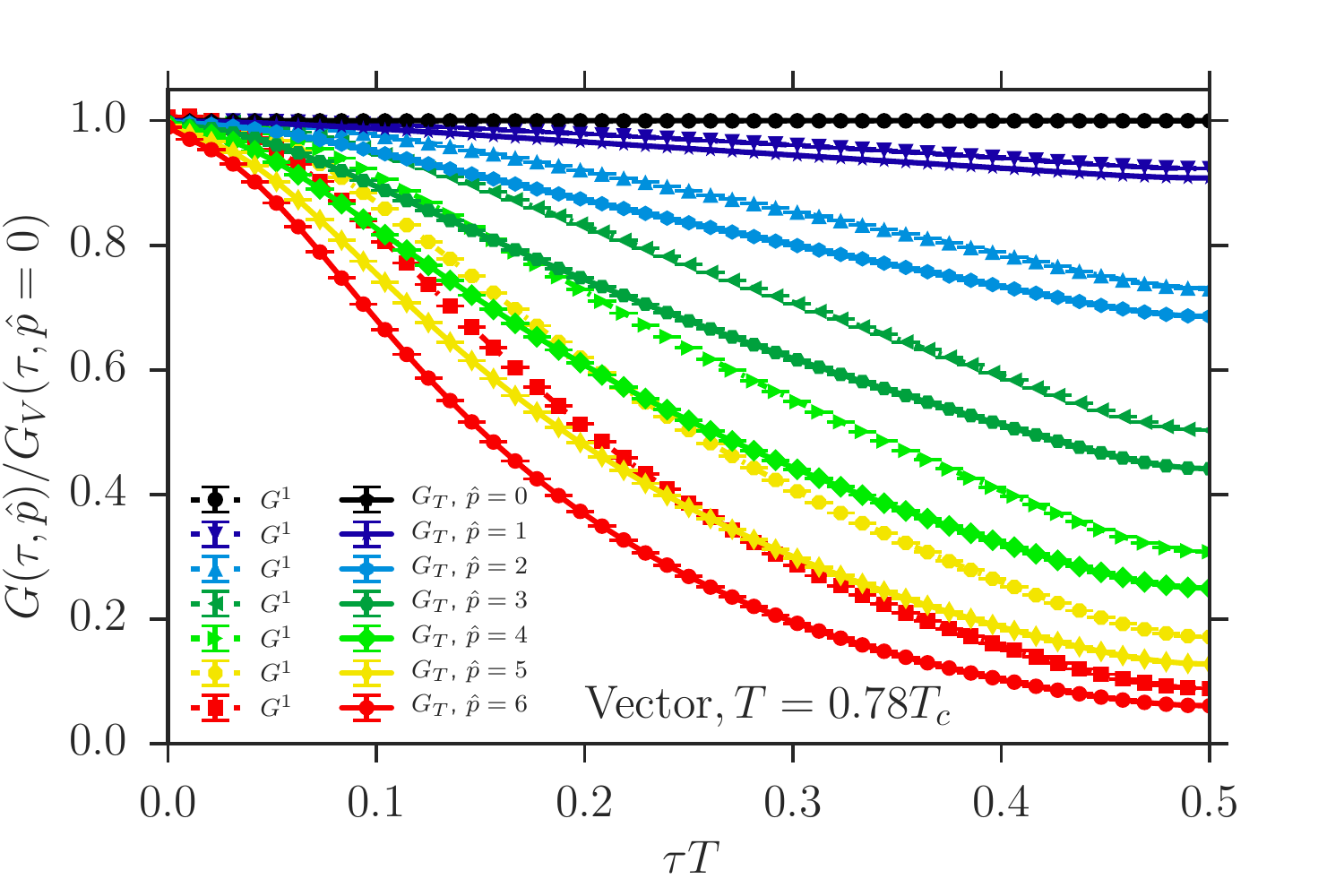}
    \includegraphics[width=0.49\textwidth]{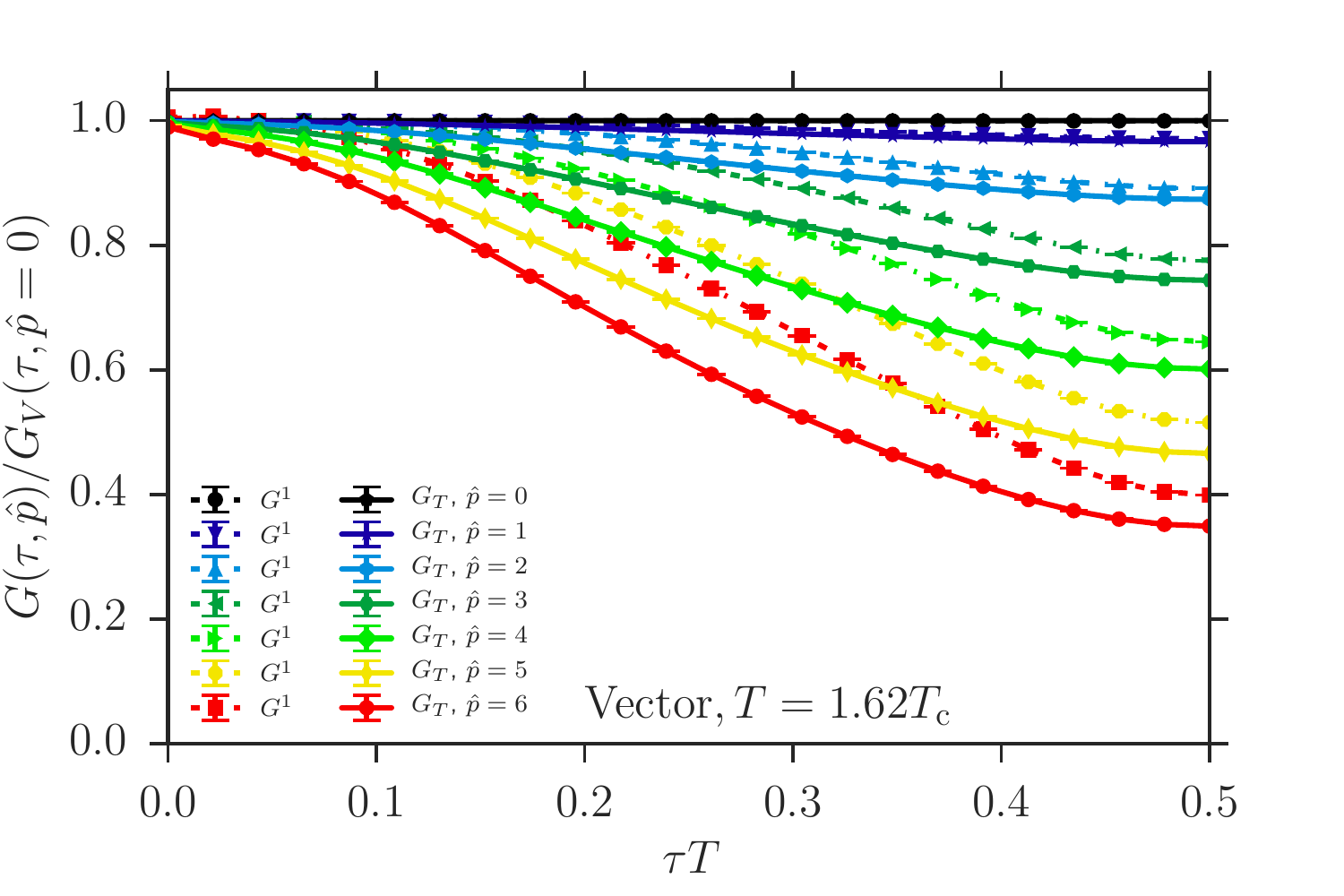}
    \caption{\label{fig:corr_hi} 
      Momentum dependence of correlation functions for the vector channel,
      $G^1(\tau,\vec{p})$ and $G_{\rm T}(\tau,{p})$,
      normalized by the correlation function with zero momentum 
      $G_{\rm V}({\tau},{0})$ for $T=0.78T_{\rm c}$ (upper) and 
      $T=1.62T_{\rm c}$ (lower).
      The dashed and solid lines represent $G^1(\tau,\vec{p})$ and
      $G_{\rm T}(\tau,{p})$, respectively.}
\end{figure}
In this section, we show the numerical results.
We first see the momentum dependence of the correlation functions.
Figure~\ref{fig:corr_hi} shows the correlation functions
$G^1(\tau,\vec{p})$ and $G_{\rm T}(\tau,p)$ 
in the vector channel normalized by those with zero momentum,
\begin{align}
  G_{\rm V}(\tau,0) =
  \frac13 \sum_{i=1,2,3} G^i(\tau,\vec{0}),
\end{align}
for various values of $\hat{p}$ below and above $T_{\rm c}$.
In the figure, the ratios $G^1(\tau,\vec{p})/G_{\rm V}(\tau,0)$ and
$G_{\rm T}(\tau,p)/G_{\rm V}(\tau,0)$ are plotted by 
the dashed and solid lines, respectively.
The errors in the figure are estimated for the ratios
$G^1(\tau,\vec{p})/G_{\rm V}(\tau,0)$ and
$G_{\rm T}(\tau,p)/G_{\rm V}(\tau,0)$ by the jackknife method;
because of the strong correlation between correlation functions 
with different $p$, these errors are suppressed compared with those 
of the correlation functions themselves.
The figure shows that the ratios
$G^1(\tau,\vec{p})/G_{\rm V}(\tau,0)$ and
$G_{\rm T}(\tau,p)/G_{\rm V}(\tau,0)$ become smaller
as $p$ is increased.
This behavior is consistent with Eqs.~(\ref{eq:correlation_function})
and (\ref{eq:peak_structure})
because as $E(p)$ becomes larger, the contribution of the bound state
to the correlation function is more suppressed.
The figure also shows that 
$G^1(\tau,\vec{p})$ and $G_{\rm T}(\tau, p)$ behave differently
even at $T=0.78T_{\rm c}$.
As momentum become larger, the separation becomes more prominent
with $G_{\rm T}(\tau, p)<G^1(\tau,\vec{p})$.
As we discussed in Sec.~\ref{sec:SPF},
$A_{\rm T}(\omega,\vec{p}) = A_{\rm L}(\omega,\vec{p}) =A_{\rm V}(P^2)$
in the vacuum. From Eq.~(\ref{eq:A_L}) we thus have
\begin{align}
  A_{\rm T}(\omega,p) = \frac{\omega^2-p^2}{\omega^2} A_1(\omega,\vec{p}).
\end{align}
Because the factor $(\omega^2-p^2)/\omega^2$ is always smaller than unity,
we have $A_{\rm T}(\omega,p) \le A_1(\omega,\vec{p})$,
which leads to $G_{\rm T}(\tau,p) \le G^1(\tau,\vec{p})$.
Similar behavior is observed for $T=1.62T_{\rm c}$.

\subsection{Spectral function}
\label{sec:spectral}

\begin{figure}[tb]
  \centering
  \includegraphics[width=0.49\textwidth]{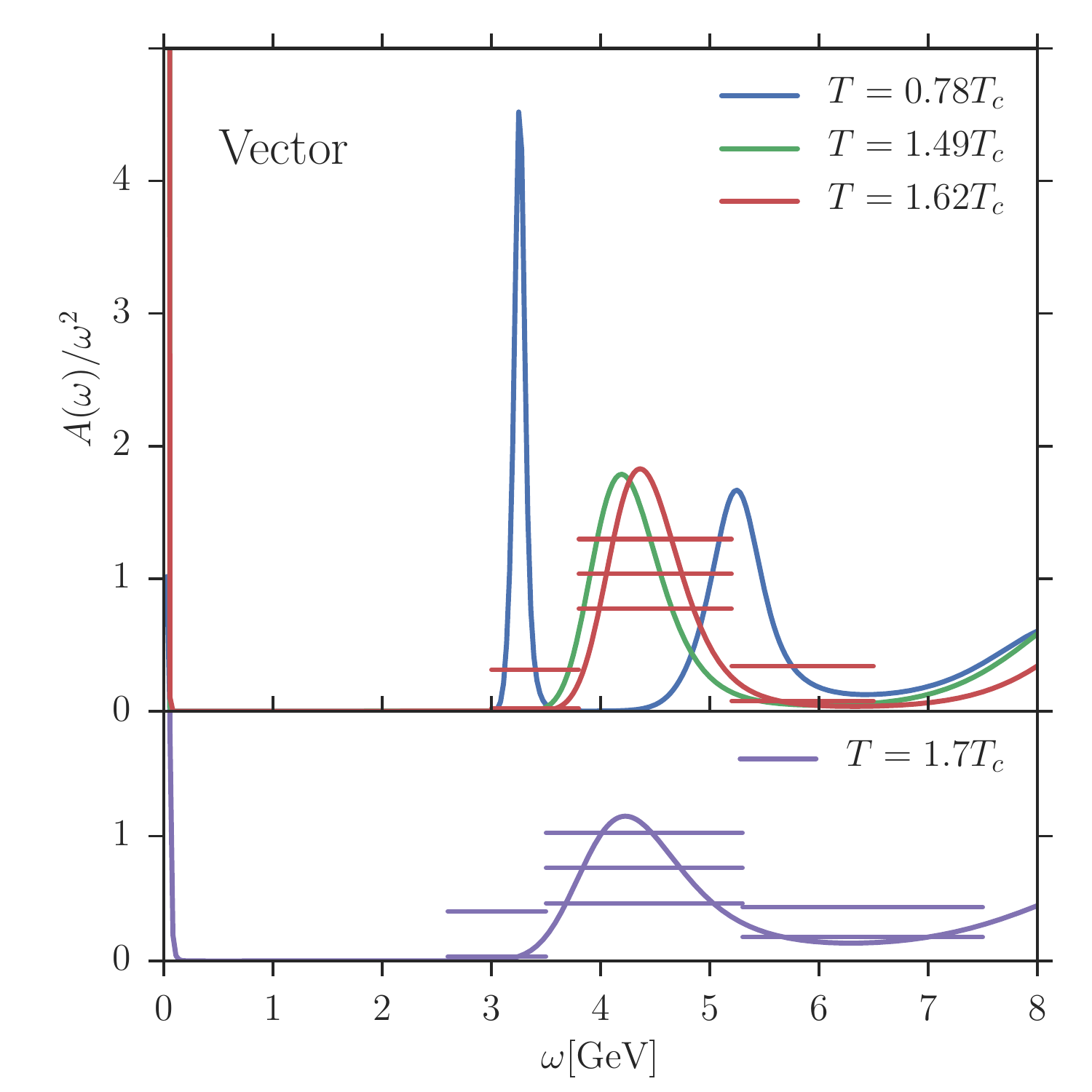}
  \includegraphics[width=0.49\textwidth]{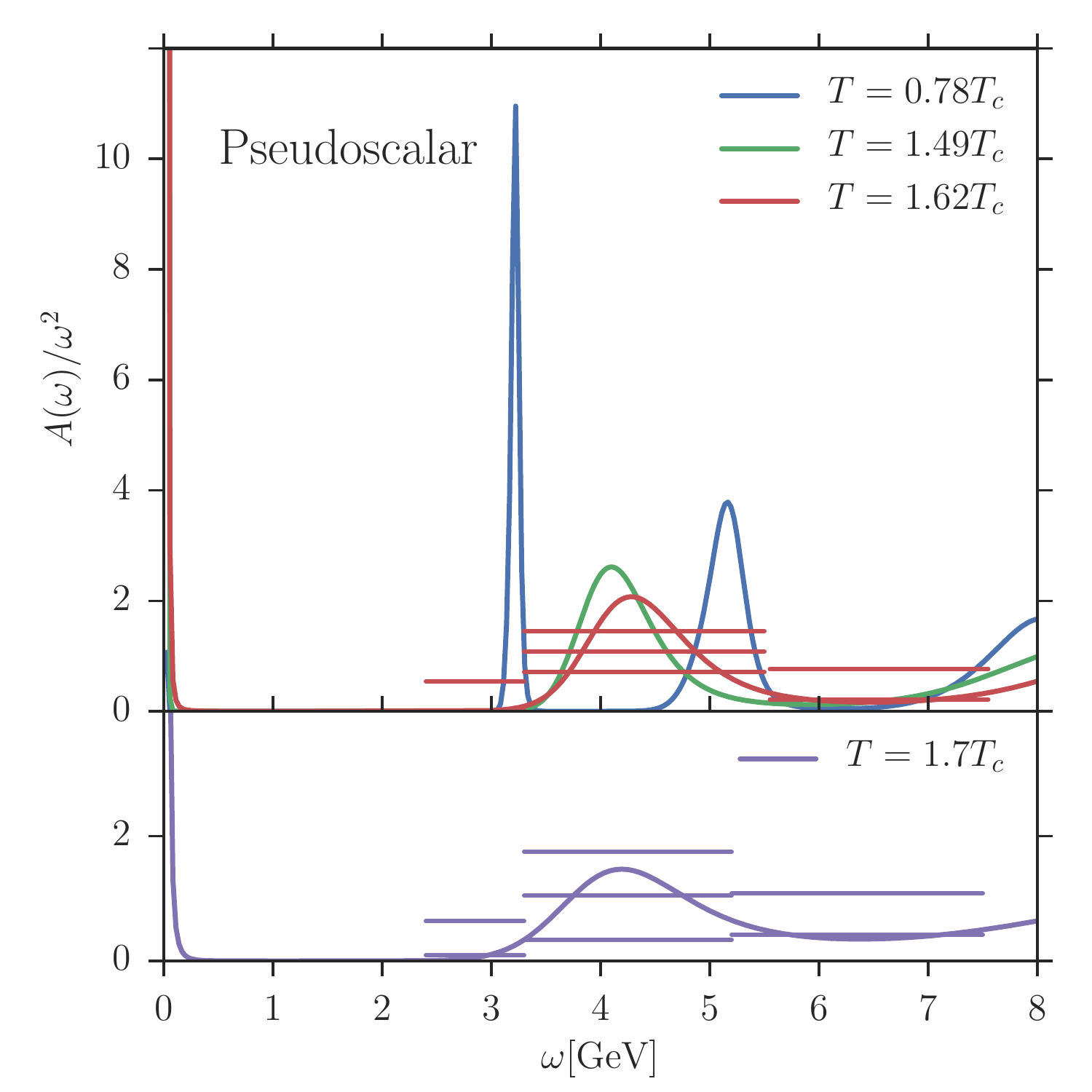}
  \caption{\label{fig:melt} 
  Spectral functions at $T/T_{\rm c}=0.78,~1.49,~1.62$, and $1.7$
  at zero momentum.
  The upper and lower panels show the vector and pseudoscalar channels,
respectively. 
  The horizontal lines show the averages of the spectral functions 
  for some ranges of $\omega$ and their errors at $T/T_{\rm c}=1.62$ and $1.70$.
}
\end{figure}

Next, we analyze the $T$ dependence of the spectral functions with MEM
and study the existence of the peaks corresponding to
the $J/\psi$ and $\eta_c$ at finite temperature.
The upper and lower panels of Fig.~\ref{fig:melt} show 
the spectral functions with zero momentum in the vector and pseudoscalar
channels, respectively.
The error bars for the average of the spectral function for some 
intervals of $\omega$ estimated by Eq.~(\ref{eq:error_A}) 
are shown by three horizontal lines for $T/T_{\rm c}=1.62$ and $1.70$.
The central lines show the averages of the spectral functions in the 
interval covered by the line, and the top and bottom ones indicate
its $1\sigma$ error band.
The result of the error analysis in the vector channel suggests
that the peak corresponding to the $J/\psi$ exists at $T=1.62T_{\rm c}$ 
with probabilistic significance.
On the other hand, for the pseudoscalar channel at $T=1.62T_{\rm c}$,
the error for the peak structure corresponding to the $\eta_c$ 
has a small overlap with the error which is put for
the right side valley of the structure.
This shows that the plausibility of the existence of the peak
is smaller than that for the vector channel. 
In other words, 
absence of the peak of $\eta_c$ at $T=1.62T_{\rm c}$ cannot be
excluded by $1\sigma$.

\begin{figure}[tb]
  \centering
  \includegraphics[width=0.49\textwidth]{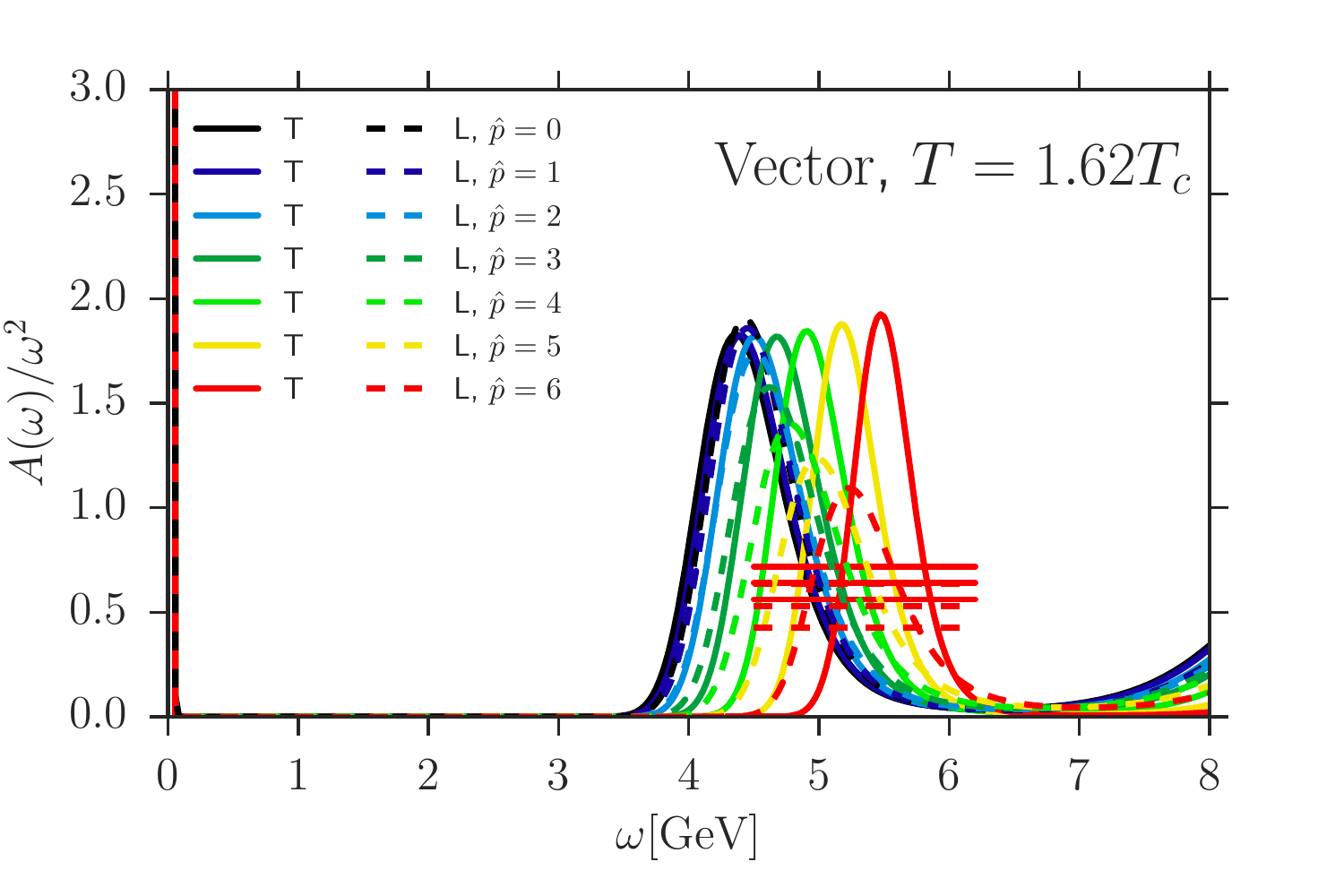}
  \includegraphics[width=0.49\textwidth]{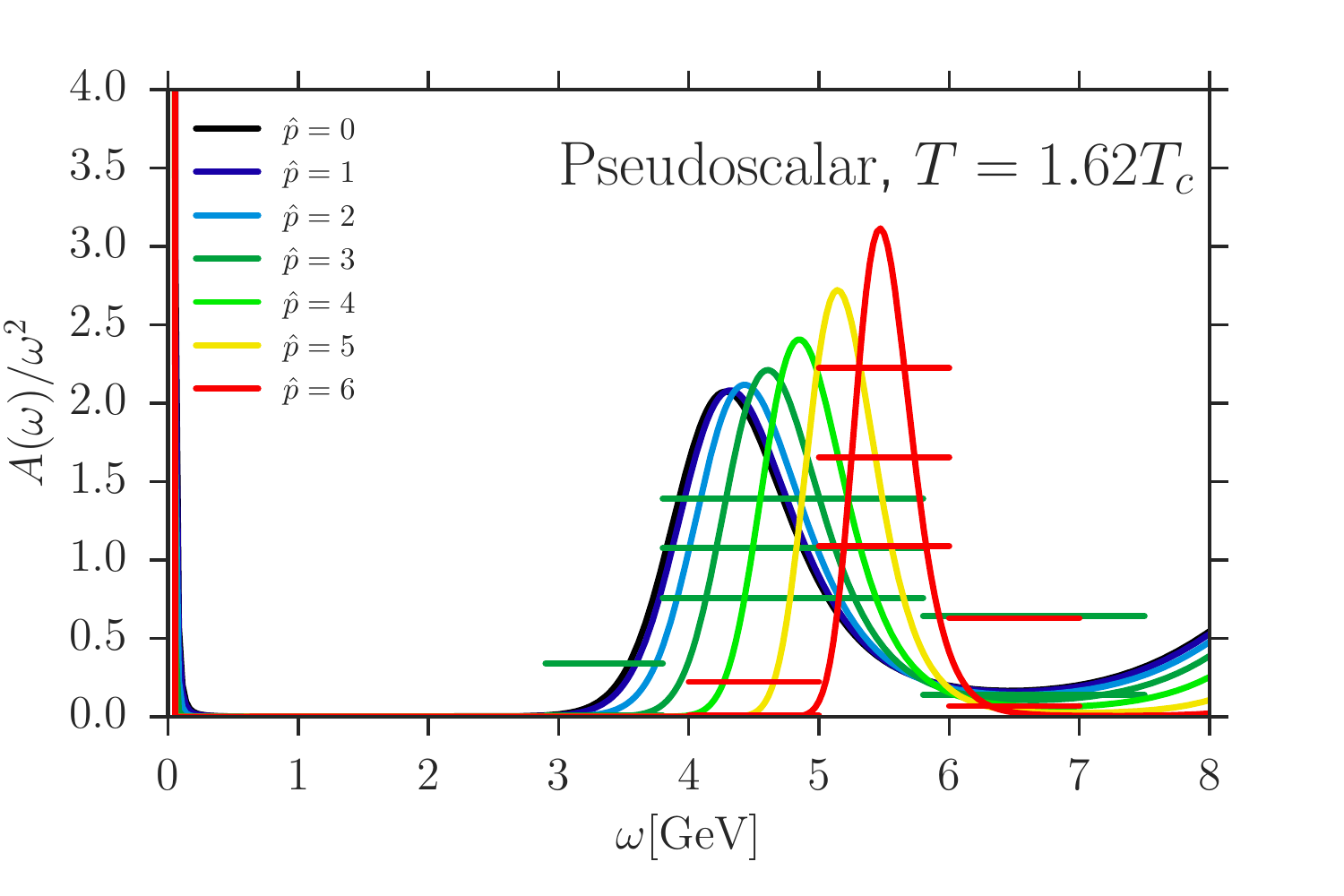}
  \caption{\label{fig:momentum_46} 
  Momentum dependence of the spectral functions at $T=1.62T_{\rm c}$.
  The upper and lower panels show the vector and pseudoscalar channels,
  respectively.
  The error bars are shown for averages of the spectral functions 
  for the vector channel at $\hat{p}=6$ and 
  for the pseudoscalar channel at $\hat{p}=3$ and $6$.
}
\end{figure}

In order to discuss the existence of the peak in the pseudoscalar channel
at $T=1.62T_{\rm c}$ and the momentum dependence of the peaks, we next show the
momentum dependence of the spectral functions in the vector and 
pseudoscalar channels at $T=1.62T_{\rm c}$ in the upper and lower 
panels in Fig.~\ref{fig:momentum_46}, respectively.
In the upper panel, $A_{\rm T}(\omega,p)$ and $A_{\rm L}(\omega,p)$ 
are shown by the solid and dashed lines, respectively.
In the lower panel, the errors for the peaks of 
the spectral functions in the pseudoscalar channel are
shown for $\hat{p}=3$ and $6$.
The lower panel suggests that the peak corresponding to the $\eta_c$
exists at $\hat{p}=3$ and $6$. 
The existence of the $J/\psi$ peak in the vector channel at nonzero 
momenta is also indicated from the upper panel.
We thus suppose that
the $J/\psi$ and $\eta_c$ survive up to $T=1.62T_{\rm c}$,
which is a consistent result 
as in previous works \cite{asakawa_j/_2004,datta_behavior_2004},
The possibility that the existence of the peak depends 
on $p$ for $T=1.62T_{\rm c}$, however, is not excluded in these analyses.

Figure~\ref{fig:momentum_46} also shows that the peaks corresponding
to the $J/\psi$ and $\eta_c$ are well isolated from the second structure
in the spectral functions.
This suggests that the dependence of $\bar{Z}(p)$ and $\bar{E}(p)$
on $\omega_{\rm max}$ is suppressed so that these quantities can 
be analyzed with small ambiguity.

\begin{figure}[tb]
  \centering
  \includegraphics[width=0.49\textwidth]{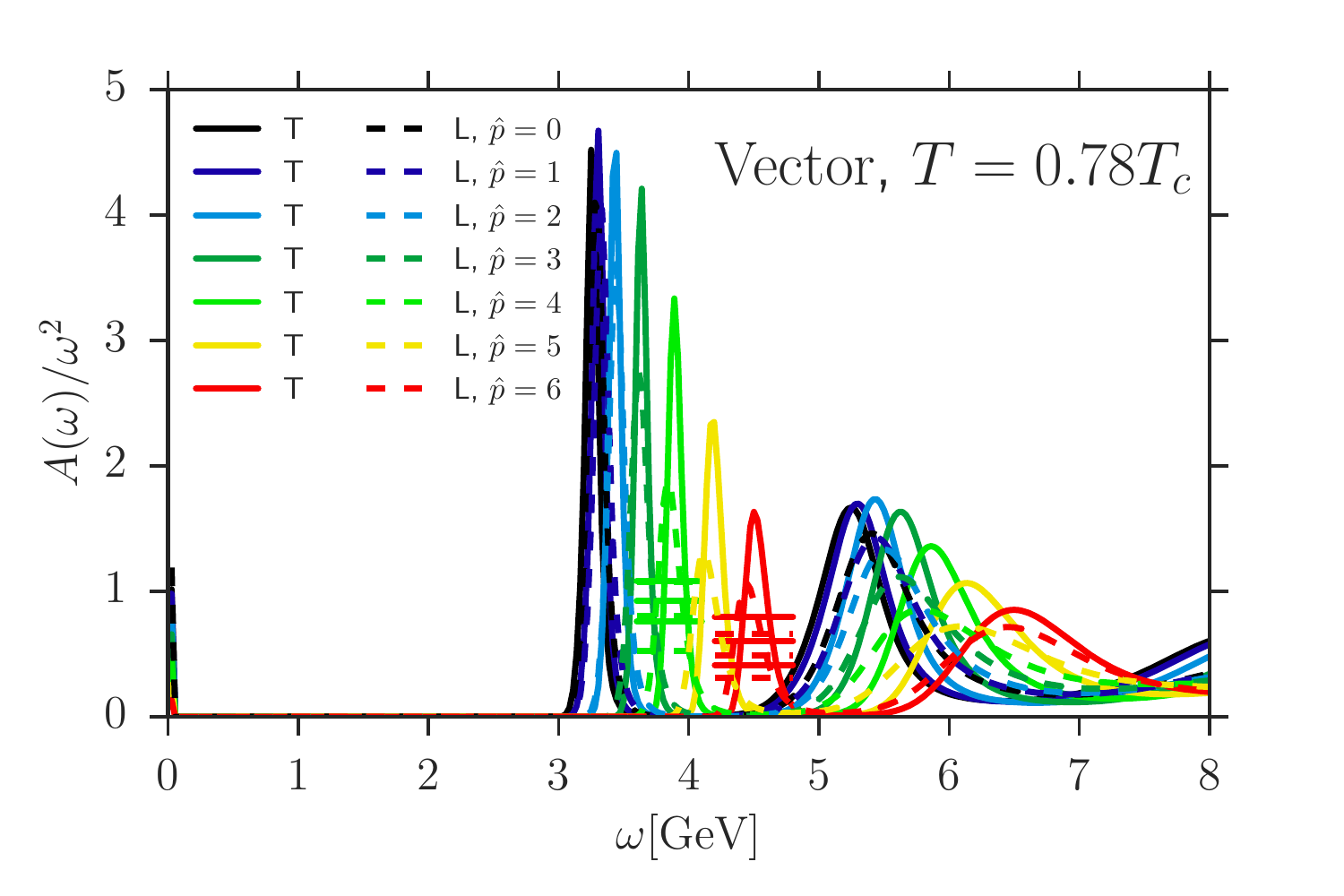}
  \caption{\label{fig:momentum_96} 
  Momentum dependence of the spectral functions
  $A_{\rm T}(\omega,p)$ and $A_{\rm L}(\omega,p)$ in the vector channel
  at $T=0.78T_{\rm c}$.
  Errors for the average of the spectral functions are shown 
  for $\hat{p}=4$ and $6$.}
\end{figure}

In Fig.~\ref{fig:momentum_96},
we show the momentum dependence of the spectral functions
in the vector channel, $A_{\rm T}(\omega,p)$ and $A_{\rm L}(\omega,p)$,
for $T=0.78T_{\rm c}$.
To see the separation of the transverse and longitudinal channels,
we show the errors for the averages of $A_{\rm T}(\omega,p)$ and 
$A_{\rm L}(\omega,p)$ with the same energy interval 
for $\hat{p}=4$ and $6$.
From the figure, one observes that the spectral functions
in the transverse and longitudinal 
channels agree with each other within the error.
This result is consistent with the vacuum property of the 
spectral functions discussed in Sec.~\ref{sec:SPF}.
It, however, is worth emphasizing that this agreement is obtained 
although $A_{\rm T}(\omega,p)$ and $A_{\rm L}(\omega,p)$ are constructed
from completely different correlation functions 
as shown in Fig.~\ref{fig:corr_hi}.
From the upper panel in Fig.~\ref{fig:momentum_46}, 
one also finds that the degeneracy of $A_{\rm T}(\omega,p)$ and 
$A_{\rm L}(\omega,p)$ is observed even for $T=1.62T_{\rm c}$.
This is a nontrivial result because these functions can behave 
differently because of the lack of Lorentz invariance.

\subsection{Residue and dispersion relation}
\label{sec:residue}

\begin{figure}[tpb]
  \centering
  \includegraphics[width=0.49\textwidth]{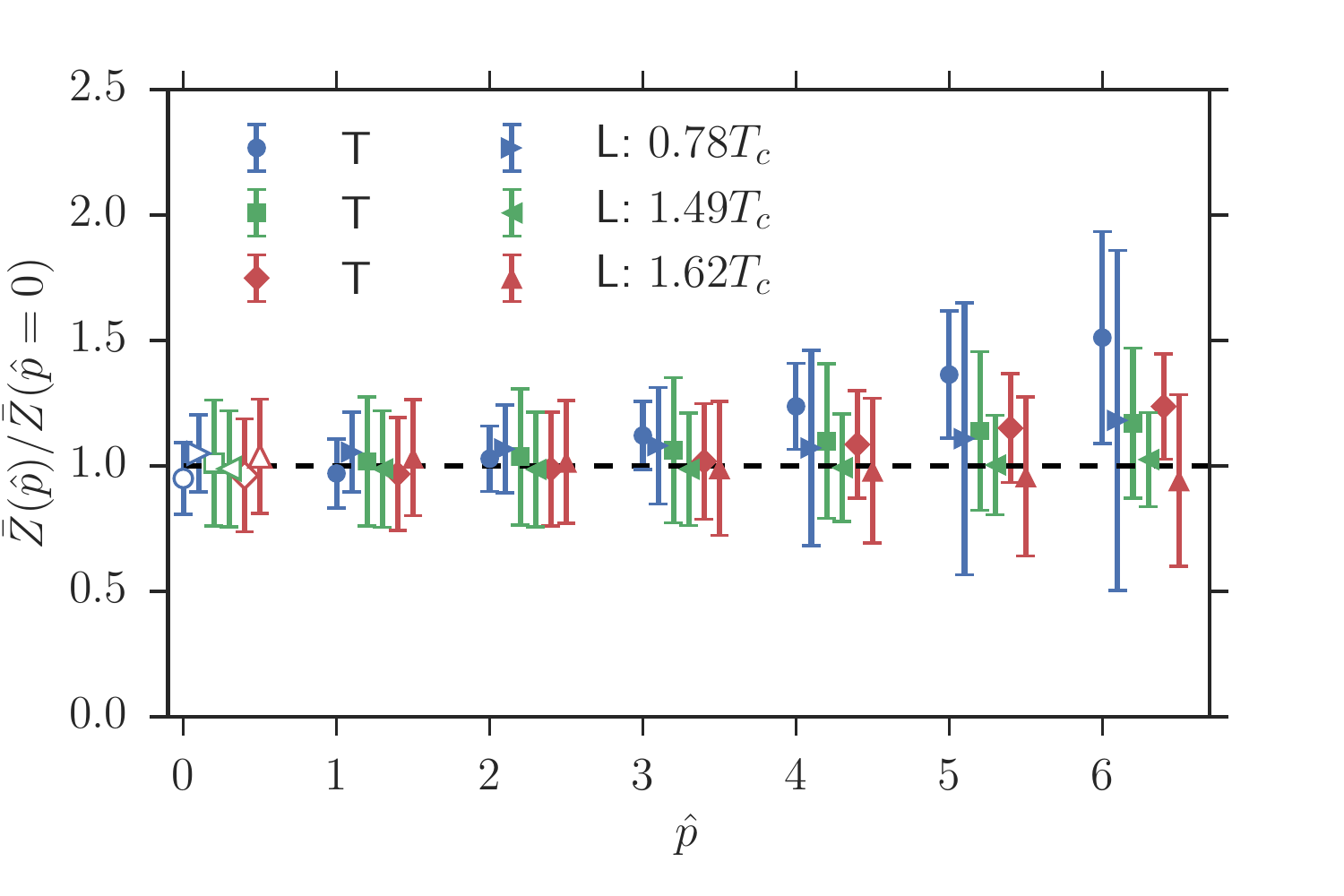}
  \includegraphics[width=0.49\textwidth]{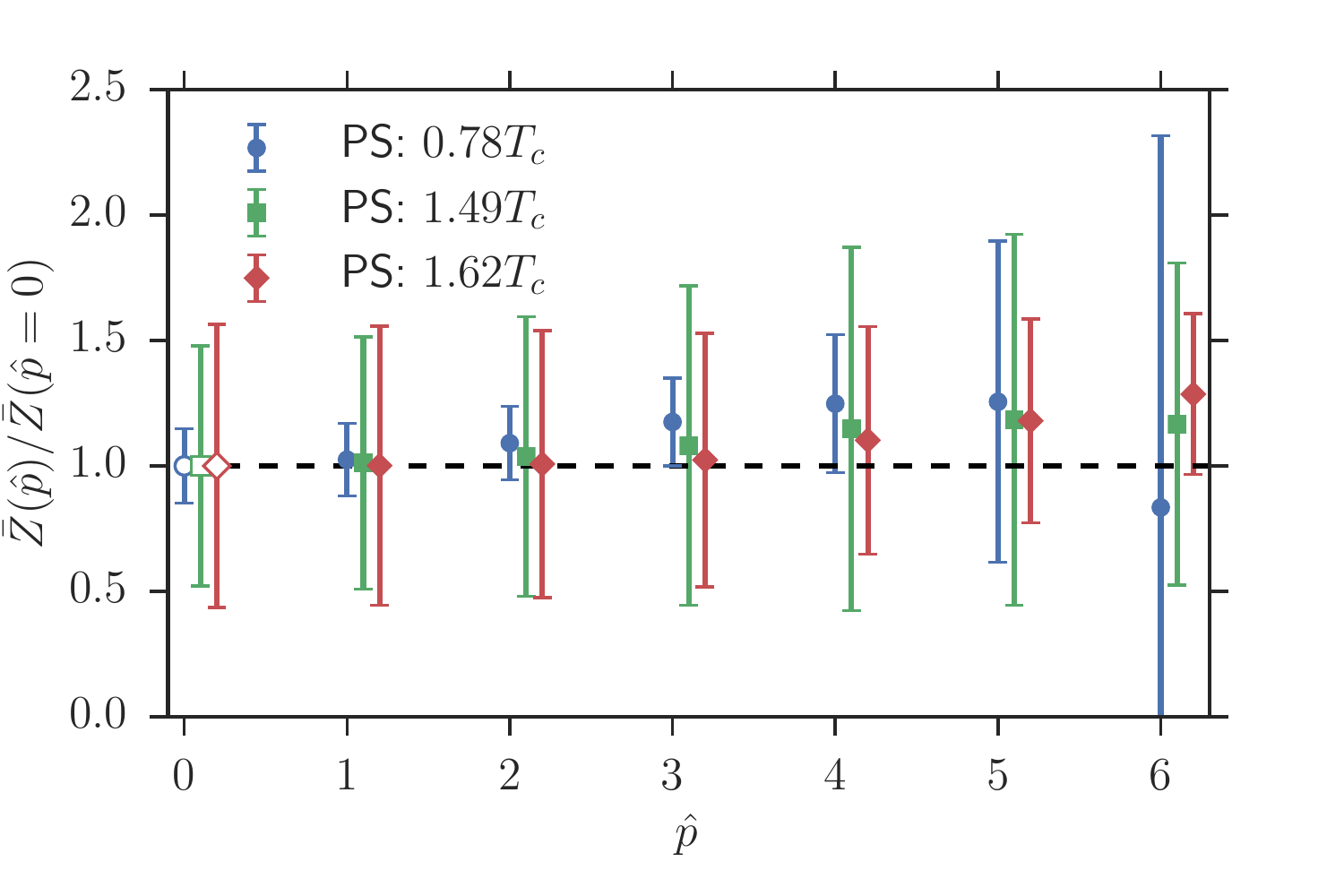}
  \caption{\label{fig:residue}
  Ratio of the residues of the peaks, $\bar{Z}(p)/\bar{Z}(0)$, 
  corresponding to the $J/\psi$ (upper) and $\eta_c$ (lower).
  %The lower figure is for the pseudoscalar channel.
  The transverse (T) and longitudinal (L) components are shown 
  in the upper panel.
}
\end{figure}

Next, we turn to $\bar{Z}(p)$ and $\bar{E}(p)$.
In Fig.~\ref{fig:residue}, we show the momentum dependence of $\bar{Z}(p)$
obtained with Eq.~(\ref{eq:barZ}) for the vector and pseudoscalar channels
for $T/T_{\rm c}=0.78,~1.49,$ and $1.62$.
In the figure, the normalized results, $\bar{Z}(p)/\bar{Z}(0)$,
are plotted in order to see the momentum dependence of $\bar{Z}(p)$.
The errors in the figure include only the one of the numerator of 
the ratio estimated by MEM.
The figure shows that $\bar{Z}(p)$ does not have 
momentum dependence within the error for all the temperatures
and all the channels for which we carried out analysis.
This result is reasonable for $T/T_{\rm c}=0.78$, at which the 
medium effects should be well suppressed.
Our analysis, however, show that $\bar{Z}(p)$ is insensitive to 
$p$ even at $T/T_{\rm c}=1.49$ and $1.62$, which is a nontrivial
result.

We note that the errors of $\bar{Z}(p)/\bar{Z}(0)$ in 
Fig.~\ref{fig:residue} would be reduced if we take into account 
the correlation between $\bar{Z}(p)$ and $\bar{Z}(0)$.
In order to estimate the correlation, however, one has to perform 
the MEM analysis for two different correlation functions
in a single analysis.
Because we perform the MEM analysis for individual momenta,
this correlation cannot be estimated in our analysis.

%Although $\bar{Z}(p)$ and $\bar{Z}(0)$ are
%strongly correlated quantities,
%one cannot estimate the error for $\bar{Z}(p)/\bar{Z}(0)$ including
%the correlation, because $\bar{Z}(p)$ and $\bar{Z}(0)$
%are estimated from the different correlation functions.
%To estimate the contribution of the correlation for the error, one has to
%perform MEM analysis for two different correlation functions
%in a single analysis.
%We are interested in comparing $\bar{Z}(0)$ and $\bar{Z}(p)$,
%therefore the errors in the figure include only the one of 
%the numerator of the ratio estimated by MEM with the method
%discussed in Sec.~\ref{sec:ZMEM}. 

\begin{figure}[tpb]
  \centering
  \includegraphics[width=0.49\textwidth]{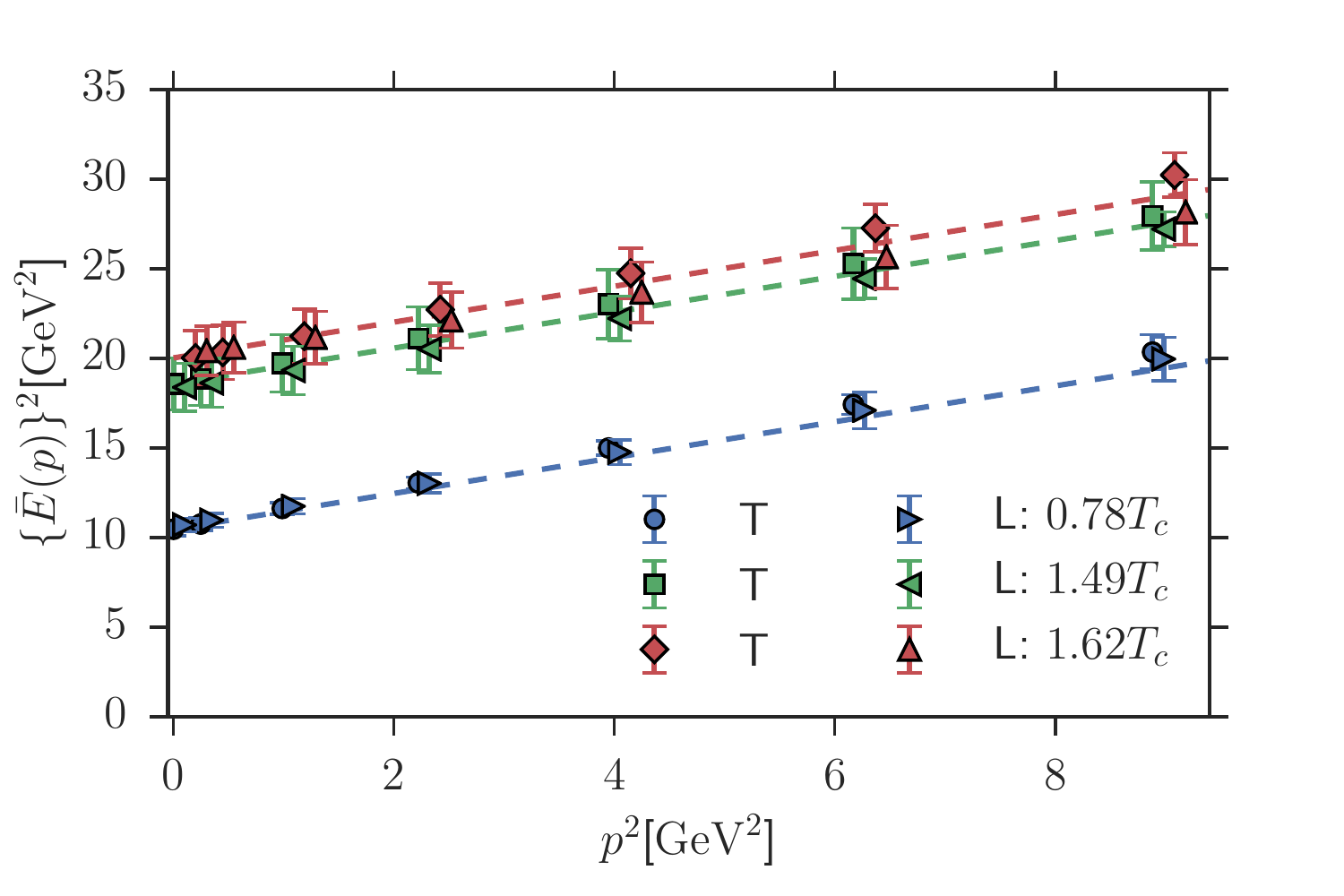}
  \includegraphics[width=0.49\textwidth]{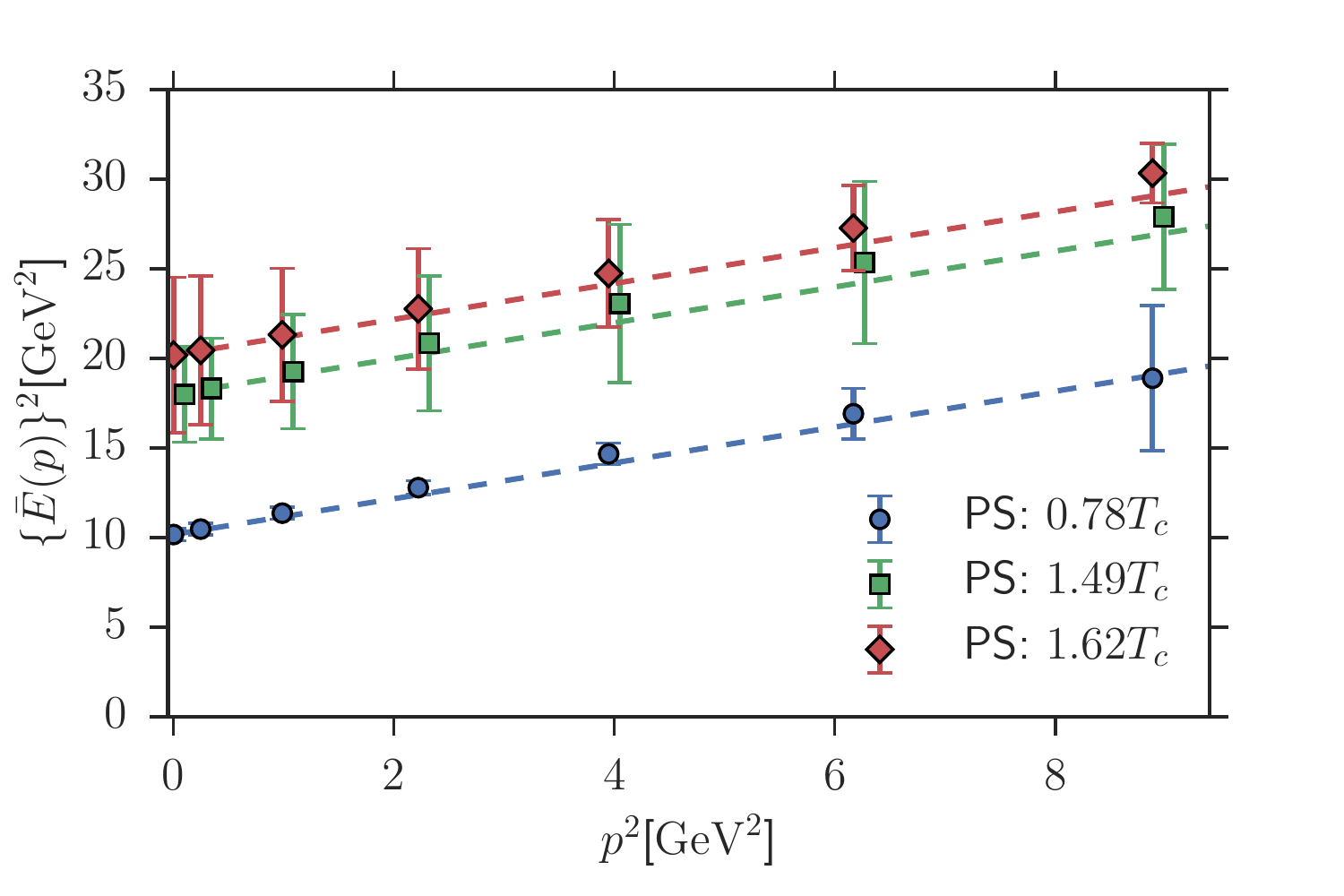}
  \caption{\label{fig:dispersion_relation} 
  Dispersion relations $\bar{E}(p)$ of the $J/\psi$ (upper) and 
  $\eta_c$ (lower) for $T/T_{\rm c}=0.78$, $1.49$, and $1.62$.
  The dashed lines indicate the vacuum dispersion relation 
  Eq.~(\ref{eq:dispersion_vacuum}) with $m=\bar{E}(0)$.
}
\end{figure}

\begin{table}
  \centering
  \begin{tabular}{c|ccc}
  \hline
  $T/T_{\rm c}$ & $0.78$    & $1.49$      &  $1.62$ \\ 
  \hline
  $J/\psi$       & $3.24(6)$ & $4.30(16)$  & $4.47(16)$ \\
  $\eta_c$       & $3.19(5)$ & $4.24(31)$  & $4.49(48)$ \\
  \hline
  \end{tabular}
  \caption{Masses of the ground states of the charmonia 
    in the vector and pseudoscalar channels defined by 
    $\bar{m}=\bar{E}(0)$.}
  \label{tab:E0}
\end{table} 

To see the medium effects on the dispersion relation,
we show the results on $\bar{E}(p)$ in Fig.~\ref{fig:dispersion_relation}.
In the figure, we plot the square of this quantity $(\bar{E}(p))^2$
as a function of $p^2$, since this plot is convenient to see the 
deviation of $\bar{E}(p)$ from the vacuum dispersion 
relation Eq.~(\ref{eq:dispersion_vacuum}).
From the figure, one first observes that the masses of the charmonia, 
defined by $\bar{m}=\bar{E}(0)$, become larger as $T$ is increased.
The values of $\bar{m}$ in the vector and pseudoscalar channels at
$T/T_{\rm c}=0.78,~1.49,$ and $1.62$ are 
listed in Table~\ref{tab:E0}.
Although such a mass shift in MEM analyses
were suggested in previous study \cite{nonaka_charmonium_2011},
our analysis confirms the medium effects on the mass of the charmonia 
with a quantitative error analysis for the first time.

In Fig.~\ref{fig:dispersion_relation}, the vacuum dispersion relation
Eq.~(\ref{eq:dispersion_vacuum}) with $m=\bar{m}$ is shown 
by the dashed lines.
The figure shows that the functional form of
$\bar{E}(p)$ is consistent with Eq.~(\ref{eq:dispersion_vacuum})
within statistics even at $T/T_{\rm c}=1.49$ and $1.62$.

\begin{figure}[tpb]
  \centering
  \includegraphics[width=0.49\textwidth]{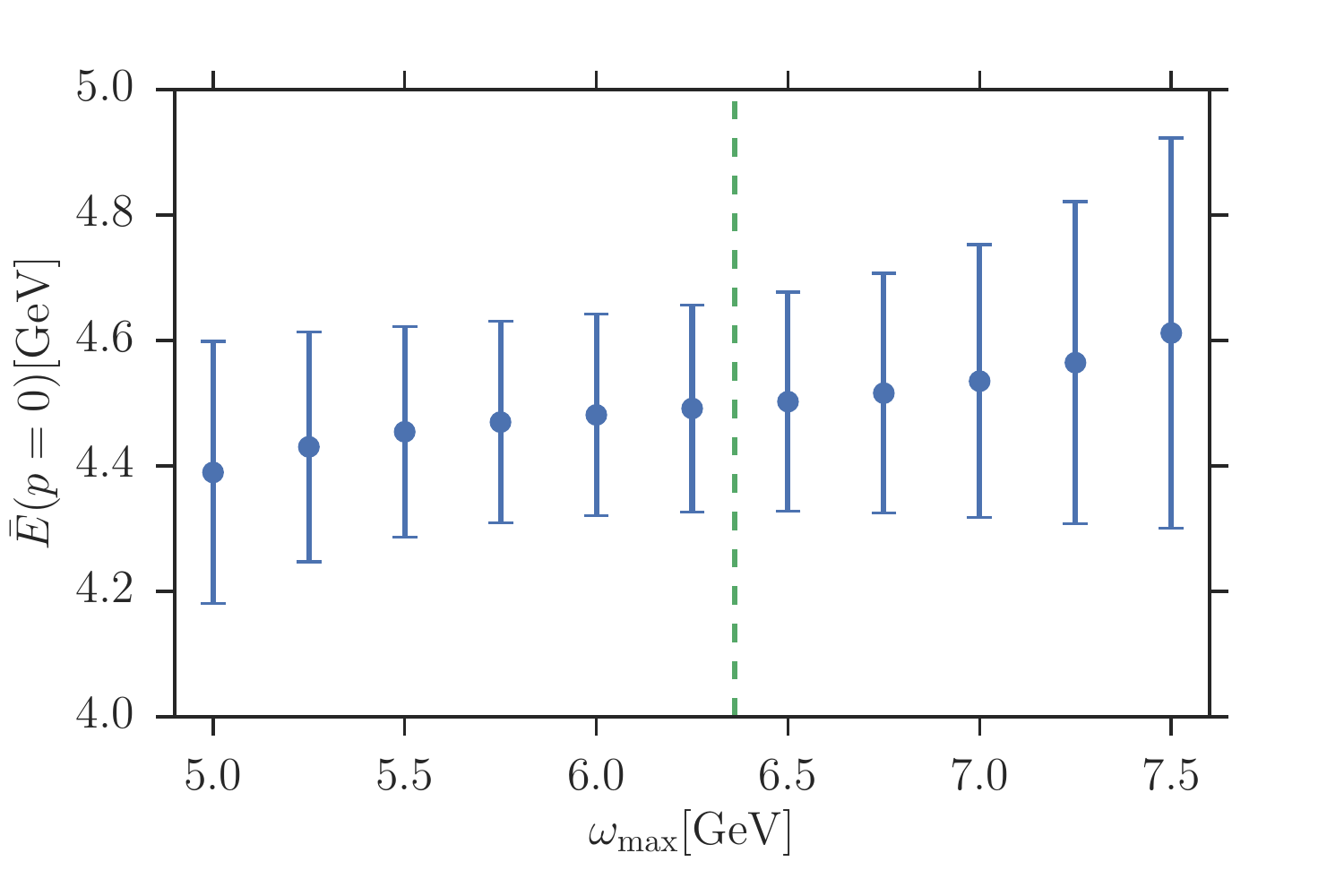}
  \caption{\label{fig:w2_dependence} 
$\omega_{\rm max}$ dependence of $\bar{E}(p=0)$ for $N_\tau=46$ 
for the $J/\psi$ peak in the vector channel.
The lower bound is fixed to $\omega_{\rm min}=3$~GeV.
The vertical dashed line shows the local minimum between the first and 
second peaks.
}
\end{figure}

In order to see the dependence of these results 
on the choice of the interval $I=[\omega_{\rm min},\omega_{\rm max}]$,
in Fig.~\ref{fig:w2_dependence}
we show the $\omega_{\rm max}$ dependence of $\bar{E}(p=0)$ for $N_\tau=46$ 
in the vector channel with $\omega_{\rm min}=3$~GeV.
The value of $\omega_{\rm max}$ used in Fig.~\ref{fig:dispersion_relation},
i.e. the minimum of the spectral function between the first and second 
peaks, is shown by the vertical dashed line.
The figure shows that the value and error of $\bar{E}(p)$ are 
insensitive to the choice of $\omega_{\rm max}$. 
In fact, the variation of the result with the change 
of $\omega_{\rm max}$ in $\pm500$~MeV is about four times smaller
than the error.
The same conclusions holds also for the other cases and 
for $\bar{Z}(p)$.
As discussed already in Sec.~\ref{sec:lattice}, the numerical 
results hardly change with the variation of $\omega_{\rm min}$. 
For example, when we choose the lower bound as $\omega_{\rm min}=2$~GeV,
the numerical result overlaps with that in 
Fig.~\ref{fig:w2_dependence} within numerical precision.
These results suggest that our analysis of $\bar{E}(p)$
is insensitive to the choice of the interval $I$ and thus is well justified.

The results in Figs.~\ref{fig:residue} and \ref{fig:dispersion_relation}
suggest that the momentum dependence of the charmonia hardly changes
from the Lorentz covariant one in Eqs.~(\ref{eq:peak_structure})
and (\ref{eq:dispersion_vacuum}) even well above $T_{\rm c}$, 
although the rest mass $m$ is significantly increased as $T$ is raised.
In vector channels, we do not observe difference between 
the transverse and longitudinal components within the error
in MEM analysis even at finite temperature.
These results are nontrivial because Lorentz symmetry is lost 
in medium, and quite interesting from the phenomenological 
points of view.

\section{Conclusion}
\label{sec:conclusion}

In this paper, we study the properties of 
charmonia at nonzero momentum in the vacuum and in medium with the lattice
Euclidean correlation functions in the pseudoscalar and vector
channels with MEM. The transverse and longitudinal components
for the vector channel are analyzed separately.
In addition to the standard analysis of spectral functions,
we focus on the residue and dispersion relations for charmonia.
To analyze these quantities with error in MEM,
we have introduced the definitions 
in Eqs.~(\ref{eq:barZ}) and (\ref{eq:barE}).
We have numerically checked that the peaks corresponding 
to the $J/\psi$ and $\eta_c$ can be studied by this analysis,
as they are well isolated and the results are insensitive to 
the choice of the interval $I$.

In the vacuum, the dispersion relations for charmonia
in all channels, pseudoscalar and vector,
are consistent with the Lorentz covariant form and 
the residues for the bound states do not show the momentum dependence.
In the vector channel, the peaks for the transverse and longitudinal
components agree with each other within probabilistic significance.
At finite temperature,
we find the significant mass enhancement of charmonia as medium effects.
On the other hand, 
the dispersion relations are consistent with 
that in the vacuum even at $T\simeq1.6T_{\rm c}$
within probabilistic significance in MEM.
Difference of the spectral functions between the
transverse and longitudinal components in the vector channel is not observed.
These results suggest an interesting observation that the medium effect
on momentum dependence is well suppressed,
although further improvement in statistics is required to obtain
more accurate conclusion.
We finally remark that these results
cannot be explained by the na\"ive potential model.
It is not explained by the threshold enhancement
\cite{mocsy_can_2008}, either. 
These results suggest that the mass shift at finite temperature
is caused by the nonperturbative interaction
between charmonia and gluons in the medium.

\begin{acknowledgments}
Numerical simulations for this study were carried out on IBM System 
Blue Gene Solution at KEK under its Large-Scale Simulation Program 
(No.~14/15-15 and 15/16-10).
The numerical analysis of this work is in part performed 
using IroIro++ code.
This work is supported in part by JSPS KAKENHI
Grant Numbers 23540307, 25800148, and 26400272.
A.~I. is supported by the Grant-in-Aid for JSPS Fellows (No.~15J01789).
\end{acknowledgments}

\appendix

\section{Failure of MEM analysis}
\label{sec:pathology}

As discussed in Sec.~\ref{sec:lattice},
in our MEM analysis we observed that the 
output spectral images behave in an unreasonable way 
in some channels and momenta on several sets of configurations.
In this appendix, we discuss this problem in detail and 
show a criterion to remove them from the analysis.

\begin{figure}[tb]
  \centering
  \includegraphics[width=0.49\textwidth]{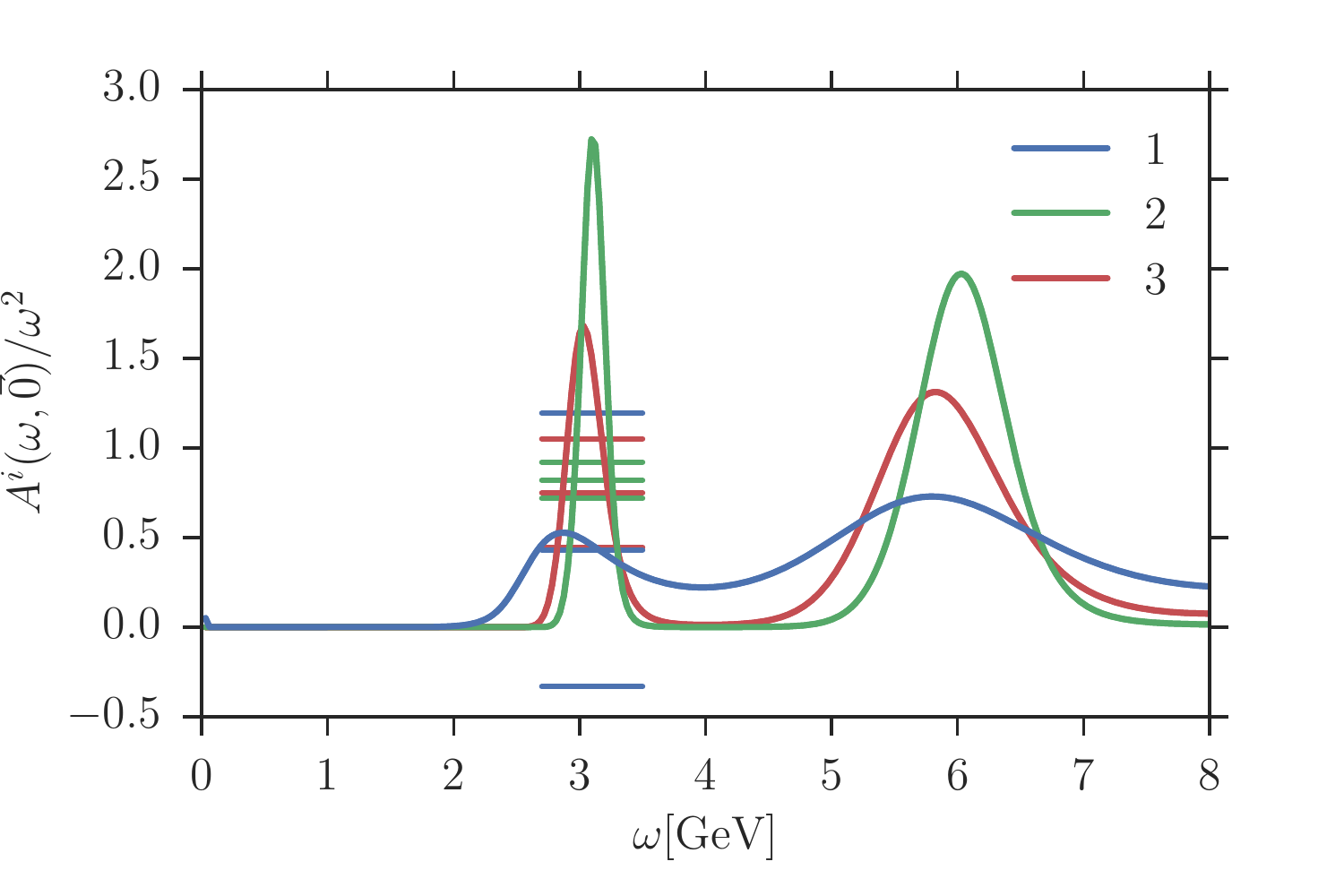}
  \caption{\label{fig:spc_54} 
  Reconstructed spectral images $A^i(\omega,\vec{0})$ with
  $i=1,~2$, and 3 and for $N_\tau=54$.
  The errors for the first peak in the spectral functions 
  are measured at the same $\omega$ interval.
}
\end{figure}

Let us first specify the problem.
In Fig.~\ref{fig:spc_54}, we show the spectral functions 
in the vector channel at zero momentum $A^i(\omega,\vec{p}=0)$
obtained by MEM for $i=1,~2$, and $3$
with $N_\tau=54$ ($T=1.38T_{\rm c}$).
Because of rotational symmetry, these three spectral functions
have to degenerate.
Moreover, since our analysis discussed in Sec.~\ref{sec:spectral}
shows the existence of a peak corresponding to the $J/\psi$
at $T/T_{\rm c}=1.49$ and $1.62$, the spectral function for 
$T=1.38T_{\rm c}$ should also have the peak.
In Fig.~\ref{fig:spc_54}, we indeed observe the peak 
in $A^2(\omega,\vec0)$ and $A^3(\omega,\vec0)$.
The peak, however, is not observed in $A^1(\omega,\vec0)$.
This result shows that the reconstruction of the spectral image
in MEM does not work well for $A^1(\omega,\vec0)$.
In the figure, the errors for the averages of $A^i(\omega,\vec0)$
around the $J/\psi$ peak are also shown.
The result shows that the averages for all channels agree
with one another within the error, although the error for 
$A^1(\omega,\vec0)$ is large.
In this sense, the MEM analysis gives the consistent results 
for these three channels.
It, however, seems obvious from the figure 
that the MEM analysis for $A^1(\omega,\vec0)$ 
is not working well compared with the other two channels.

We observed this kind of unstable results in some channels and 
momenta for $N_\tau=48$ and $54$.
We have checked that the increase of statistics does not 
always resolve this problem; this problem sometimes manifests itself 
when the number of gauge configurations is increased.
We have also checked that this problem does not come from 
the finite numerical precision in our MEM code 
by confirming that the same problem shows up even if we 
change the numerical precision 
in our code from double to quadruple.
It has been also checked that the change of the default model
does not cure this problem.

\begin{figure}[tpb]
  \centering
  \includegraphics[width=0.49\textwidth]{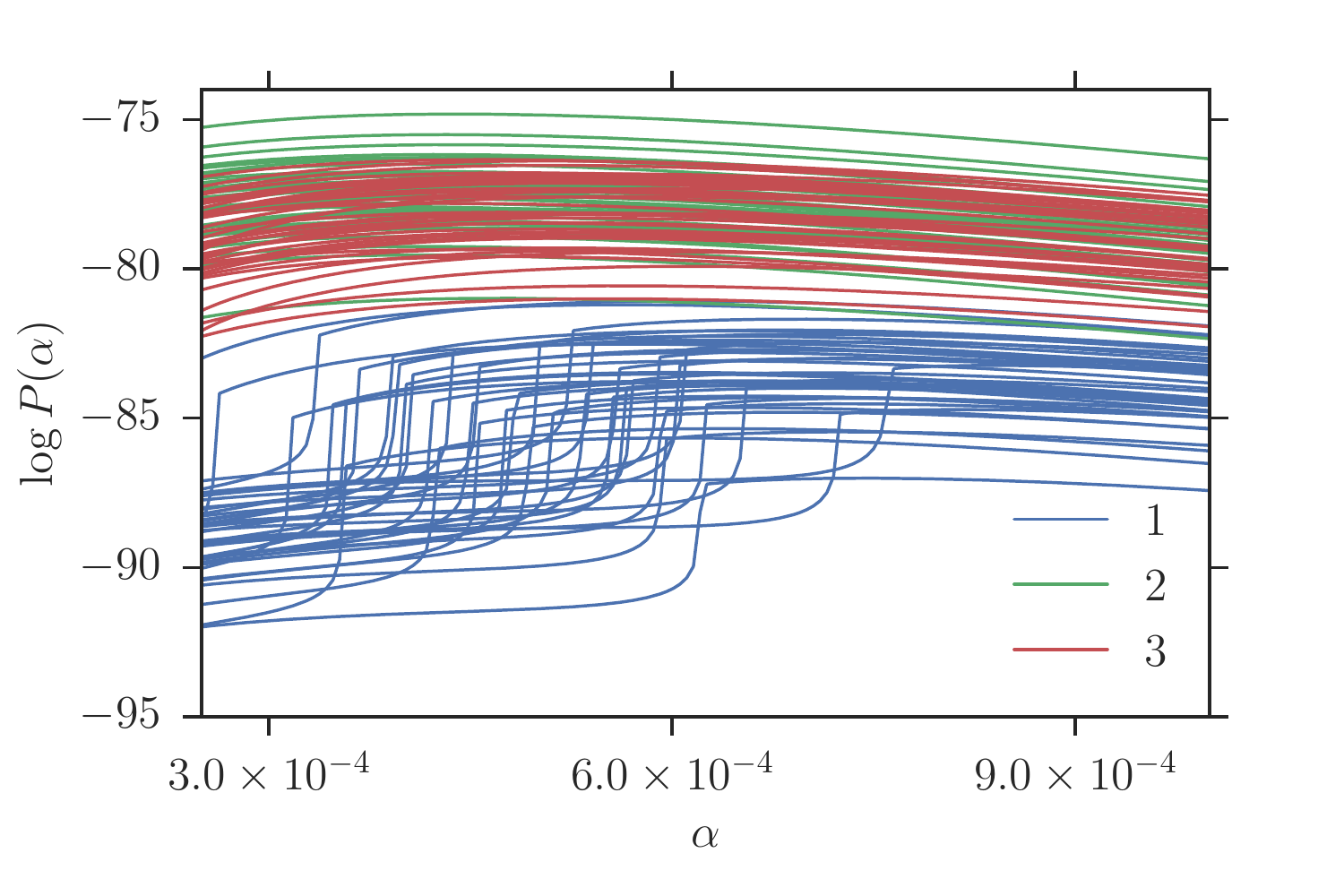}
  \caption{\label{fig:p_alpha_54} 
  $\alpha$ dependence of $P^i(\alpha)$ with $i=1,2,$ and $3$.
  The results are shown for $25$ jackknife samples for each $i$.
  The reference point of the vertical axis is arbitrary.
}
\end{figure}

We found that when the output spectral image shows an unstable behavior,
the probability $P(\alpha)$ in Eq.~(\ref{eq:P(a)}) behaves 
pathologically as a function of $\alpha$.
In Fig.~\ref{fig:p_alpha_54}, we show the probabilities $P^i(\alpha)$ 
for the three channels corresponding to the results in 
Fig.~\ref{fig:spc_54}.
In the figure, $P^i(\alpha)$ is shown for $25$ jackknife samples,
i.e. the results for $25$ sets of configurations in which $1/25$ 
succeeding configurations are removed from the total configurations.
From the figure, one finds that $P^1(\alpha)$ has an almost discontinuous 
kink structure while $P^2(\alpha)$ and $P^3(\alpha)$ behave smoothly
as a function of $\alpha$.
The figure also shows that the existence of the kink is robust
against the small variation of the set of gauge configurations.
We found that when $P(\alpha)$ has such kink structures,
the output spectral image behaves in an unstable way as in 
$A^1(\omega,\vec{p})$ in Fig.~\ref{fig:spc_54}.

At present, we have not clarified the origin of this pathological
behaviors.
One possibility for the origin of this behavior is the numerical precision
of the lattice simulation and correlation functions.
The numerical simulations, however, have been performed in double 
precision and it is difficult to alter the precision.

The pathological behavior of $P(\alpha)$ is observed on the analysis
for $N_\tau=54$ and $48$, while we do not observe 
it for $N_\tau=96,~50,~46$, and $44$.
In our study, we simply exclude the results of $N_\tau=54$ and 
$48$ from our analysis and concentrate on the other four $N_\tau$ values,
which do not have the problem.

%\bibliography{ref_trans3}
\input{./mem_v4.bbl}
\end{document}